\documentclass[twocolumn]{aastex631}

\usepackage{times}
\usepackage{amsmath,amssymb,amstext}
\hypersetup{linkcolor=red,citecolor=blue,filecolor=cyan,urlcolor=blue}
\usepackage{tabularx}
\usepackage{longtable}
\usepackage{float}
\usepackage{natbib}
\usepackage{graphicx,color}
\usepackage{txfonts}

\usepackage{multirow}
\usepackage{array}
\usepackage{rotating}
\usepackage{sidecap}
\usepackage{hyperref}
\usepackage{epstopdf}
\usepackage{footnote}
\usepackage{tabularx}
\usepackage{booktabs}
\usepackage{longtable}
\usepackage{tabu}
\usepackage{longtable}

\newcommand{\kms}{km\,s$^{-1}$}

\begin{document}
\title{{\large{\bf Solar Eruption Onset and Particle Acceleration in Nested-Null Topologies}}}

\author{Pankaj Kumar\altaffiliation{1,2}}

\affiliation{Department of Physics, American University, Washington, DC 20016, USA}
\affiliation{Heliophysics Science Division, NASA Goddard Space Flight Center, Greenbelt, MD, 20771, USA}

\author{Judith T.\ Karpen}
\affiliation{Heliophysics Science Division, NASA Goddard Space Flight Center, Greenbelt, MD, 20771, USA}

\author{Peter F.\ Wyper}
\affiliation{Department of Mathematical Sciences, Durham University, Durham DH1 3LE, UK}

\author{David Lario}
\affiliation{Heliophysics Science Division, NASA Goddard Space Flight Center, Greenbelt, MD, 20771, USA}

\author{Spiro K. Antiochos}
\affiliation{Heliophysics Science Division, NASA Goddard Space Flight Center, Greenbelt, MD, 20771, USA}

\author{C.\ Richard DeVore}
\affiliation{Retired Independent Consultant}
\affiliation{Heliophysics Science Division, NASA Goddard Space Flight Center, Greenbelt, MD, 20771, USA}

\email{pankaj.kumar@nasa.gov}

\begin{abstract}
The magnetic breakout model explains a variety of solar eruptions, ranging from small-scale jets to large-scale coronal mass ejections (CMEs). Most of our previous studies are focused on jets and CMEs in single null-point topologies. Here, we investigate the initiation of CMEs and associated particle acceleration in a double null-point (or nested fan-spine) topology during multiple homologous M- and X-class flares from an active region. The initiation of the flare and associated eruption begins with inflow structures moving towards the inner null of the closed fan-spine topology. Simultaneous slow flare reconnection below a small filament formed a hot flux rope along with expansion of the overlying flux during slow breakout reconnection at the inner null. The first explosive breakout reconnection of the flux rope at the inner null produced a circular and remote ribbons along with successful eruption of the flux rope and associated fast EUV (shock) wave. Simultaneous flare reconnection beneath the erupting flux rope produced a typical two-ribbon flare along with two hard X-ray footpoint sources. When the flux rope (with shock) reaches the outer null, the second explosive breakout reconnection produces another large-scale remote ribbon. The radio observations reveal quasiperiodic Type III bursts (period=100-s) and a Type II burst during the breakout reconnection near the inner and outer nulls, along with gradual solar energetic particles (SEPs) observed at 1 AU for magnetically connected events. This study highlight the importance of two successive breakout reconnection in the initiation of CMEs in nested-null topologies and associated particle acceleration/release into the interplanetary medium. 
 The particles are accelerated by the shock ahead of the flux rope, which formed during the inner breakout reconnection. These findings have significant implications for particle acceleration and escape processes in multi-scale null-point topologies that produce jets and CMEs.
\end{abstract}

\section{INTRODUCTION}\label{intro}
Coronal mass ejections (CMEs) are large-scale eruptions of magnetized plasma from the Sun that play a crucial role in space weather by driving shocks, generating solar energetic particles (SEPs), and causing geomagnetic disturbances at Earth \citep{gopal2006,chen2011,vorlidas2020}. The initiation and evolution of CMEs are governed by the restructuring of the solar magnetic field, often involving breakout reconnection \citep{antiochos1999}. 
The breakout model has been extensively studied in single null-point topologies, where a multipolar field configuration facilitates reconnection at a null point, weakening the overlying magnetic constraints and enabling the flux-rope eruption \citep{antiochos1999, lynch2008, karpen2012, lynch2013, guidoni2016, wyper2021, wyper2024}.

Coronal jets \citep[see review by][]{raouafi2016}, often observed in bright points or small active regions, are frequently associated with filament eruptions \citep{sterling2015}. MHD models indicate that these jets are produced by breakout reconnection when the flux rope interacts with the breakout current sheet (BCS) formed near a null point \citep{wyper2017,wyper2018}. Observations further suggest that nearly all jets occur in small-scale null-point topologies, where breakout (or interchange) reconnection of the flux rope occurs near the null point  \citep{kumar2018,kumar2019b,kumar2019a}. While flare reconnection beneath the filament builds the flux rope, breakout reconnection destroys it, generating straight or helical coronal jets.

In the case of large-scale eruptions in pseudostreamers \citep{wang2007}, a similar breakout reconnection process can occur at the null point \citep{lynch2013}. The same fundamental mechanism that drives coronal jets in small bright points --- filament-channel eruption followed by breakout reconnection --- can operate on a larger scale, leading to more substantial eruptions, i.e., narrow jets to hybrid/medium and wide CMEs (breakout continuum), and associated fast EUV waves \citep{kumar2021}. Observations suggest that most CMEs from pseudostreamers are narrow and slow, classifying them as a distinct class of CMEs \citep{wang2015,wang2018,wang2023}, although there are a few instances where pseudostreamers can produce fast and wide CMEs accompanied by shocks \citep{kumar2021}. Additionally, recent observations indicate that magnetic reconnection at the BCS in small null-point topologies can produce prolonged quasiperiodic energy release and jets, which are relevant for understanding the formation of solar wind microstreams and switchbacks \citep{kumar2023a,kumar2024a}. This implies that pseudostreamer eruptions also serve as a large-scale counterpart to similar interchange/breakout reconnection processes as in small-scale jets, contributing mass flux to the solar wind \citep{kumar2022a,raouafi2023} and dynamic events in the heliosphere \citep{kumar2021,wyper2024,lynch2025}.

Recent studies show that more complex magnetic topologies, such as nested-null configurations, can also facilitate solar eruptions through successive breakout reconnection episodes \citep{karpen2024,yao2024}.
 Unlike single-null configurations, double-null systems introduce additional reconnection sites where energy release can occur in multiple stages. Recently, we reported multiwavelength observations of a failed eruption in a nested-null topology \citep{karpen2024}. A flux rope formed through flare reconnection and successfully erupted via breakout reconnection at the inner null, but was unable to escape from the pseudostreamer. No CME was observed in the coronagraph images; instead, only a weak shock propagating laterally outward was detected. Because the flux rope rotated in the low corona, we concluded that the rotation of the flux rope likely altered its magnetic-field direction during the eruption. Consequently, no significant reconnection occurred at the outer null between the flux rope and the open field surrounding the pseudostreamer, leading to the stalled eruption.

The key questions that we explore in this paper are: (i) How do eruptions in nested-null topologies differ from their single fan-spine counterparts?
(ii) Why do some eruptions fail while others succeed? (iii) Where and how are particles energized and accelerated? This paper presents a detailed analysis of multiple homologous M- and X-class flares and their associated eruptions within one nested-null topology during three days, based on EUV, X-ray, radio, and in-situ particle observations.
Previous studies have focused on different aspects of these homologous events, such as EIT/Moreton waves and the energy build-up/release process \citep[e.g., ][]{long2019,sahu2023}. \citet{liu2015} focused only on the inner fan-spine structure and the associated X-class flare, without considering the breakout model.

Understanding particle acceleration in such topologies is crucial for linking magnetic reconnection processes to SEP events. Current sheets formed at stressed nulls may serve as efficient particle accelerators, e.g., where energetic electrons escape along open field lines during breakout reconnection, producing recurrent Type III radio bursts \citep{kumar2016,kumar2017,chen2018,Pallister2021}. MHD simulations indicate that flare-accelerated particles are typically trapped within the flux rope but can escape into the interplanetary medium through interchange reconnection \citep{masson2013,masson2019}. CME-driven shocks formed in the low corona during the eruption also can accelerate particles, generating Type II radio bursts and gradual SEPs \citep[e.g., ][]{reames1999,gopalswamy2005,lario2016}. 

We analyzed multiwavelength observations of homologous eruptions from a nested-null topology on March 28, 29 and 30, 2014. The observations analyzed here provide new insights into the role of successive breakout reconnection episodes in particle acceleration. All events (E1, E2, E3, E4) were associated with filament eruptions, accompanied by initial breakout reconnection at the inner closed fan-spine topology, and produced moderate-speed CMEs along with shocks in the lower corona. We observed a second breakout reconnection when the flux rope and shock encountered the outer null. Both breakout intervals generated strong radio bursts (Types II and III). Moreover, SEPs were detected at 1 AU by the Solar and Heliospheric Observatory (SOHO) and Wind spacecrafts during the eruptions (E1,E3,E4), which were magnetically connected to those spacecrafts.   


\begin{figure*}
\centering{
\includegraphics[width=18cm]{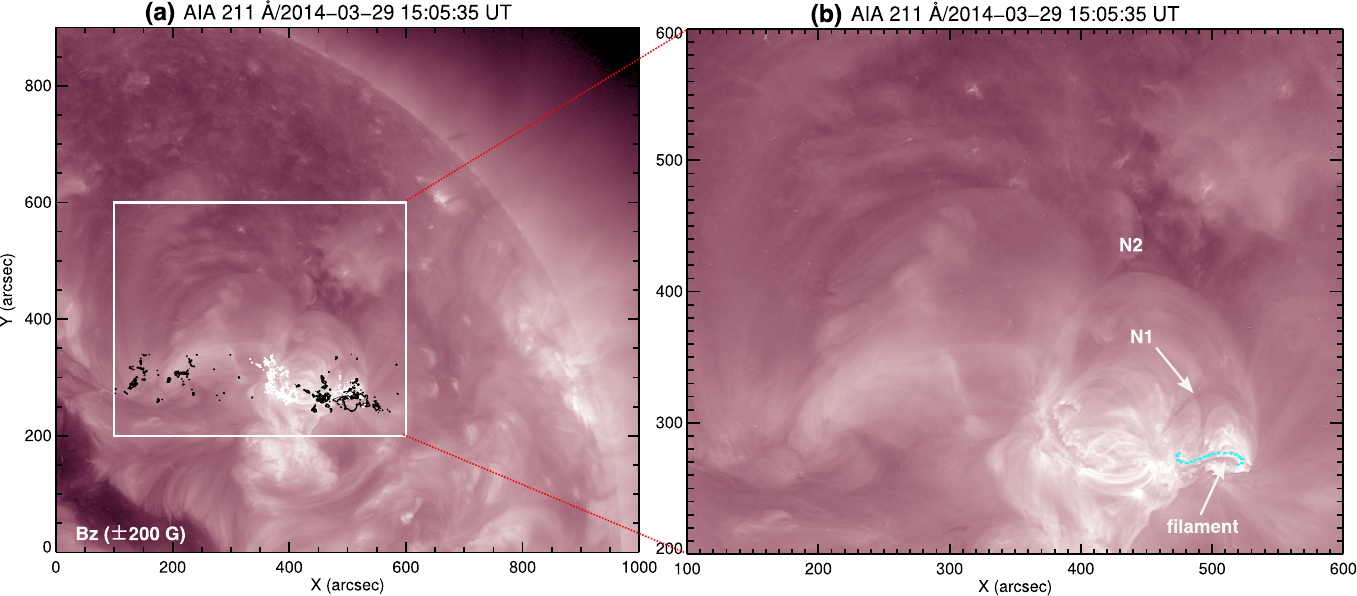}
\includegraphics[width=17cm]{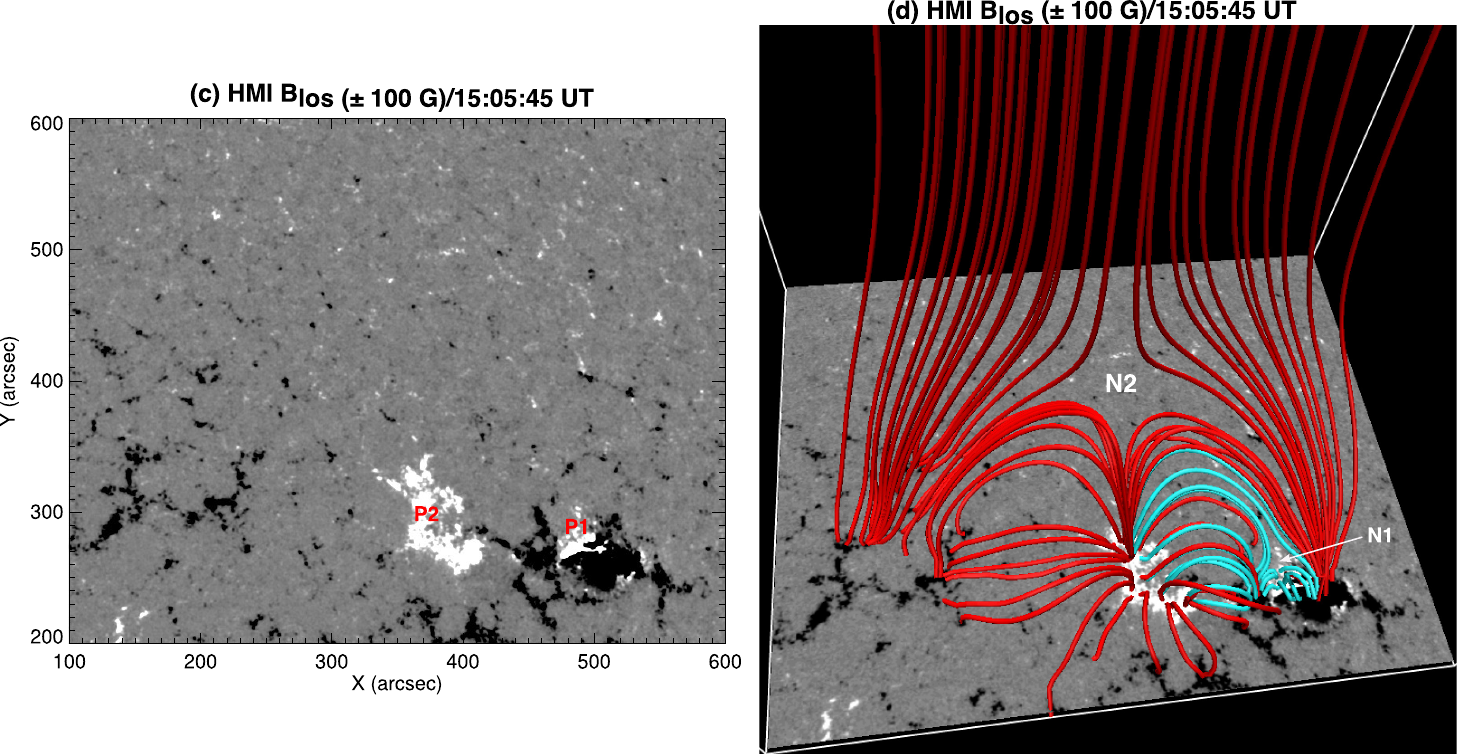}
}
\caption{{\bf Plasma and magnetic configuration of AR NOAA 12017 on  March 29, 2014.} (a) AIA 211~{\AA} image of the active region overlaid by HMI magnetogram contours ($\pm$200 G). White (black) contours indicate positive (negative) polarities. (b) The zoomed view of the region outlined  by the white box in panel (a). N1 represents the approximate position of the lower (inner) null of the nested-null topology, while  N2 indicates the approximate position of the higher (outer) null. The arrow points to an small S-shaped filament (outlined by the cyan dashed line). (c) HMI magnetogram (scaled between $\pm$100 G) at 15:06:45 UT. P1 and P2 represent the minority-polarity fluxes (positive) surrounded by opposite polarities. (d) Potential field extrapolation of the active region utilizing the magnetogram shown in panel (c). The cyan field lines illustrate the inner fan-spine topology rooted within the pseudostreamer and surrounding open field (red). N1 and N2 represent the inner and outer nulls, respectively.} 
\label{fig1}
\end{figure*}

\section{DATA}\label{obs}
We analyzed {\it Solar Dynamics Observatory} (SDO)/Atmospheric Image Assembly (AIA; \citealt{lemen2012}) full-disk images of the Sun (field-of-view $\approx$1.3~R$_\odot$) with a spatial resolution of 1.5$\arcsec$ (0.6$\arcsec$~pixel$^{-1}$) and a cadence of 12~s, in the following channels: 304~\AA\ (\ion{He}{2}, at temperature $T\approx 0.05$~MK), 171~\AA\ (\ion{Fe}{9}, $T\approx 0.7$~MK), 211~\AA\ (\ion{Fe}{14}, $T\approx 2$~MK), AIA 94~\AA\ (\ion{Fe}{10}, \ion{Fe}{18}, $T\approx$1 MK, $T\approx$6.3 MK), and 131~\AA\ (\ion{Fe}{8}, \ion{Fe}{21}, \ion{Fe}{23}, i.e., 0.4, 10, 16 MK) images.
 A 3D noise-gating technique \citep{deforest2017} was used to clean the SDO AIA images.
 We used photospheric magnetograms from the Helioseismic and Magnetic Imager \citep[HMI;][]{scherrer2012} onboard SDO to analyze the magnetic topology of the source region. To determine the magnetic configuration, we applied a potential field extrapolation code \citep{nakagawa1972} available in the SolarSoft SSWIDL GX simulator package \citep{nita2015}.

The Extreme Ultraviolet Imager \citep[EUVI;][]{Wuelser2004,Howard2008} on {\rm Solar TErrestrial RElations Observatory} Behind (STEREO-B) observed the studied flares behind the east limb. The longitudinal separation between SDO and STEREO-A (STEREO-B) was 154.6$^{\circ}$(-163.5$^{\circ}$) on March 29, 2014. We used STEREO-B COR1 (1.3-4 $R_{\sun}$) and STEREO-A COR2 (2-15 $R_{\sun}$) \citep{thompson2003} and SOHO Large Angle and Spectrometric COronograph (LASCO) C2 (2-6 $R_{\sun}$) \citep{brueckner1995,yashiro2004} images for CME propagation in the interplanetary medium. 

To investigate the inner electron-acceleration sites, we used hard X-ray lightcurves and images from the Reuven Ramaty High Energy Solar Spectroscopic Imager (RHESSI; \citet{Lin2002}), reconstructed with the PIXON algorithm \citep{Metcalf1996,Aschwanden2004}. We chose an integration time of 24 s for the image reconstruction in different energy channels (6-12, 12-25, 25-50, 50-100, 100-300 keV) using detectors 3-9. 

We used radio imaging at metric and decimetric frequencies (150–450 MHz) from the Nan{\c{c}}ay Radioheliograph (NRH) observations \citep{kerdraon1997}. 
We utilized dynamic radio spectra obtained by the Radio Solar Telescope Network (RSTN) Learmonth Radio Observatory, e-Callisto (extended-Compound Astronomical Low-frequency Low-cost Instrument for Spectroscopy and Transportable Observatory:\citealt{benz2009}) for metric/decimetric emission from the low corona, and Wind/WAVES \citep{bougeret1995} for the interplanetary medium.

We used Wind 3DP (Three-Dimensional Plasma and Energetic Particles: \citealt{lin1999}), SOHO EPHIN (Electron Proton Helium Instrument: \citealt{muller1995}), and SOHO ERNE (Energetic and Relativistic Nuclei and Electron: \citealt{torsti1995}) observations for the SEPs (electrons and protons) at 1 AU.

\begin{figure*}
\centering{
\includegraphics[width=18cm]{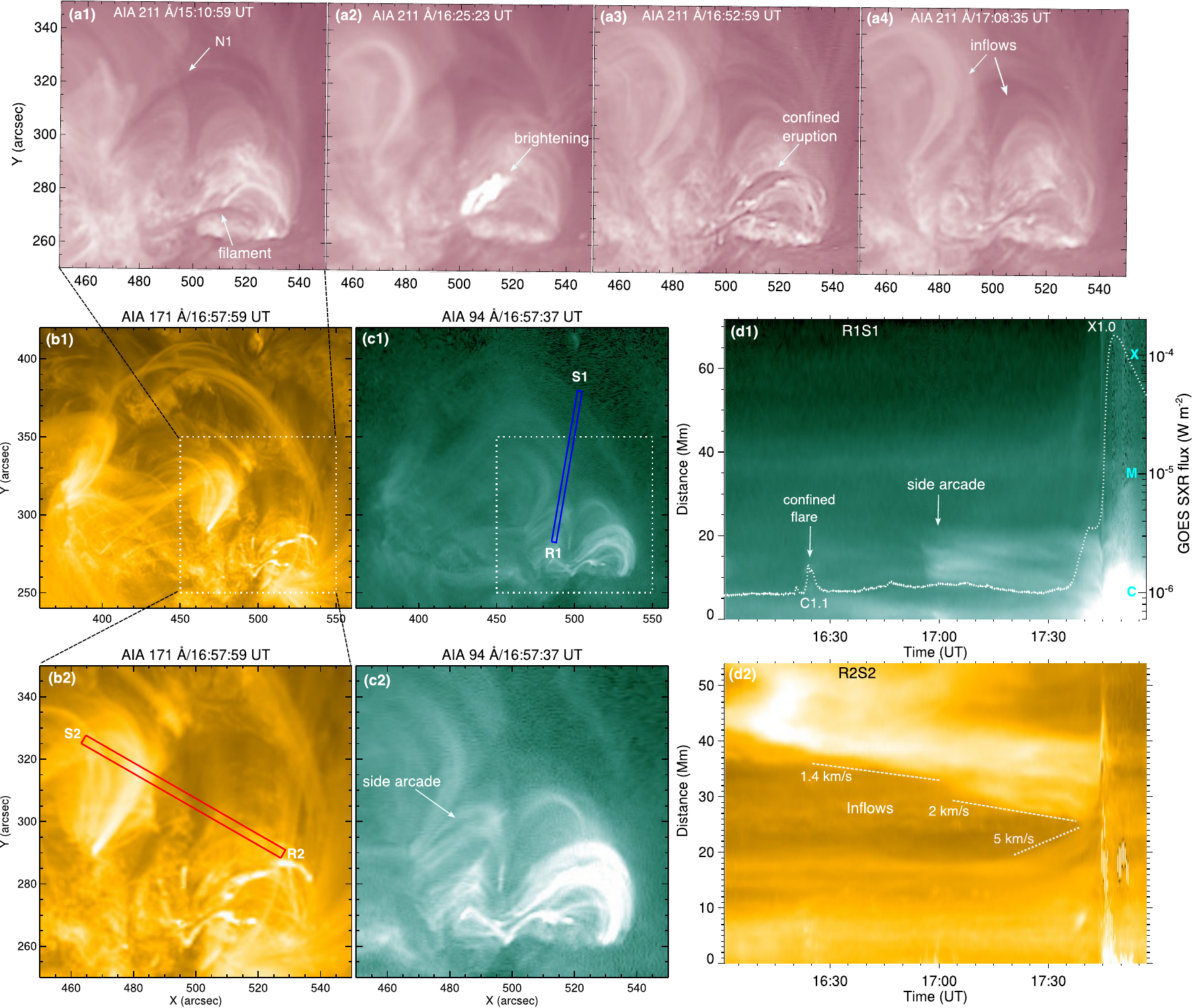}
}
\caption{{\bf Pre-eruption activities prior to Event E1 on March 29, 2014.} (a1–a4) AIA 211~{\AA} images showing a confined filament eruption and an associated C1.1 flare within the inner fan-spine system. N1 denotes the inner null (X-point). (b1, b2, c1, c2) AIA 171~{\AA} and 94~{\AA} images capturing the pre-eruption phases. R1S1 and R2S2 indicate the slices used to generate the time-distance intensity maps. (d1, d2) Time-distance intensity maps along slices R1S1 and R2S2 during 16:00-18:00 UT. The white dotted line in (d1) displays the GOES soft X-ray flux profile in the 1–8~{\AA} channel (scale on the right Y-axis).} 
\label{fig1a}
\end{figure*}
\begin{figure*}
\centering{
\includegraphics[width=18cm]{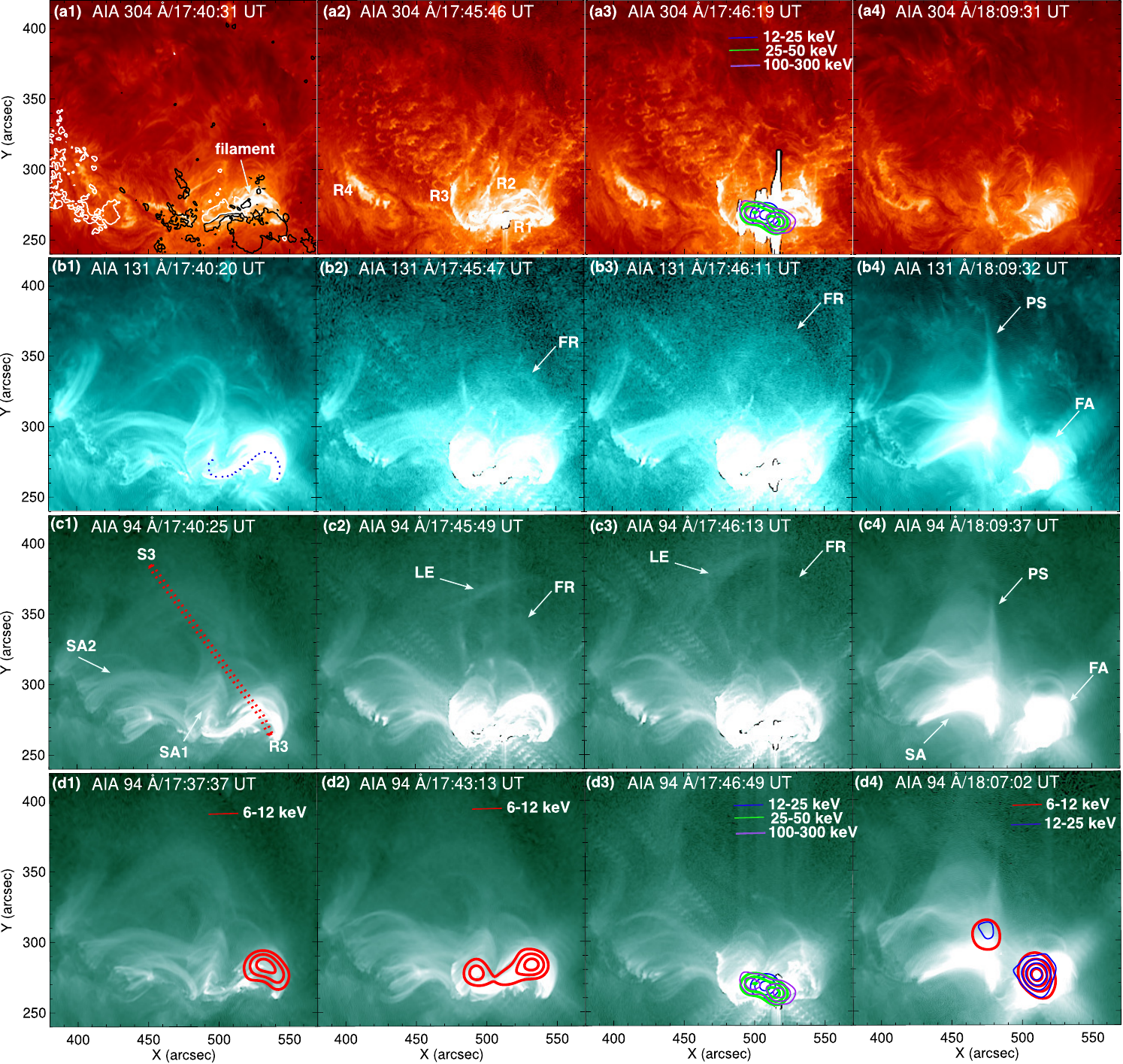}
}
\caption{{\bf Formation and eruption of the flux rope and associated X1.0 flare for Event E1.} (a1–a4) AIA 304~{\AA} images showing a filament and multiple ribbons (R1, R2, R3, R4) prior to and during the eruption. The first image is overlaid with HMI magnetogram contours (scale =$\pm$200 G) representing positive (white) and negative (black) polarities. RHESSI contours (12–25 keV, 25–50 keV, and 100–300 keV) are overlaid on the AIA 304~{\AA} image in panel (a3), with contour levels at 50$\%$, 70$\%$, and 90$\%$ of the peak intensity.
(b1–b4, c1–c4) AIA 131~{\AA} and 94~{\AA} images showing the flux rope (FR), leading edge (LE), and associated plasma sheet (PS). FA represents the flare arcade, while SA1 and SA2 indicate side arcades. R3S3 marks a slice used to generate the time-distance intensity map in Figure \ref{fig1c}(d1).
(d1–d4) RHESSI contours (6-12 keV, 12-25 keV, 25-50 keV, 100-300 keV) overlaid on AIA 94~{\AA} images during the pre-eruption, eruption, and post-eruption phases. The contour levels in panels (d1–d3) are 50$\%$, 70$\%$, and 90$\%$ of the peak intensity, whereas in panel (d4), the contour levels are 30$\%$, 50$\%$, 70$\%$, and 90$\%$ of the peak intensity.} 
\label{fig1b}
\end{figure*}

\begin{figure*}
\centering{
\includegraphics[width=18cm]{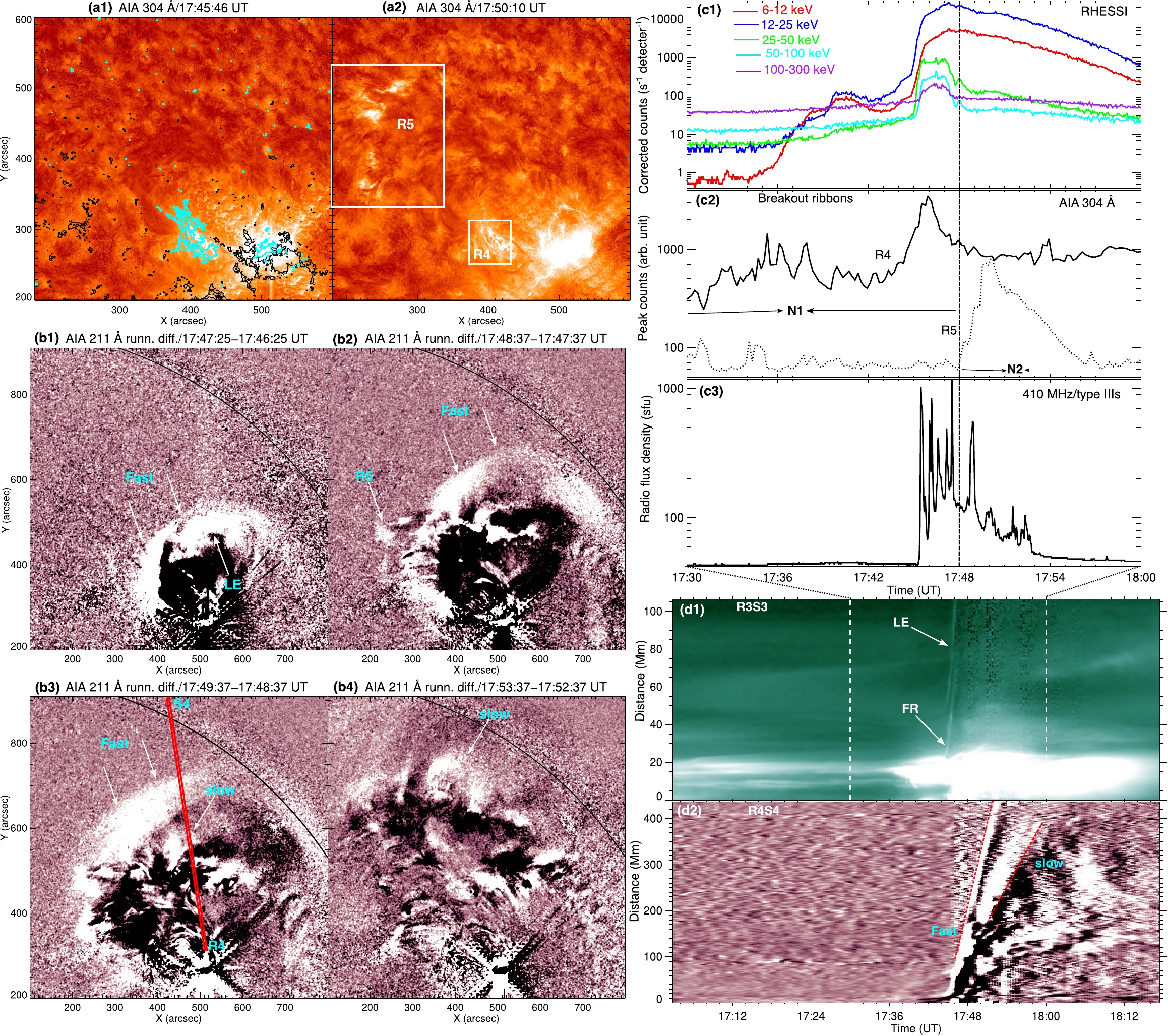}
}
\caption{{\bf Kinematics of the flux rope and associated fast/slow EUV waves during Event E1.}
(a1, a2) AIA 304~{\AA} images showing remote ribbons R4 and R5 during the flux rope eruption. The first image is overlaid with HMI magnetogram contours (scale = $\pm$100 G), representing positive (cyan) and negative (black) polarities.
(b1–b4) AIA 211~{\AA} running-difference images capturing the fast and slow EUV waves during the eruption. LE denotes the leading edge of the flux rope. The solid black curve traces the solar limb.
(c1) RHESSI X-ray flux profiles (4-s cadence) in five different energy channels.
(c2) AIA 304~{\AA} peak counts (12-s cadence) extracted from ribbons R4 and R5 within the boxes outlined in (a2). The vertical dashed line indicates the onset of energy release near N2. The time intervals of energy release at N1 and N2 are indicated by arrows. 
(c3) RSTN radio flux density profile (410 MHz, 1-s cadence) from the Sagamore Hill Solar Observatory.
(d1, d2) Time-distance intensity plots along slices R3S3 (see Figure \ref{fig1b}(c1)) and R4S4 (see panel (b3)), derived from AIA 94~{\AA} and 211~{\AA} images.
} 
\label{fig1c}
\end{figure*}
\begin{figure*}
\centering{
\includegraphics[width=18cm]{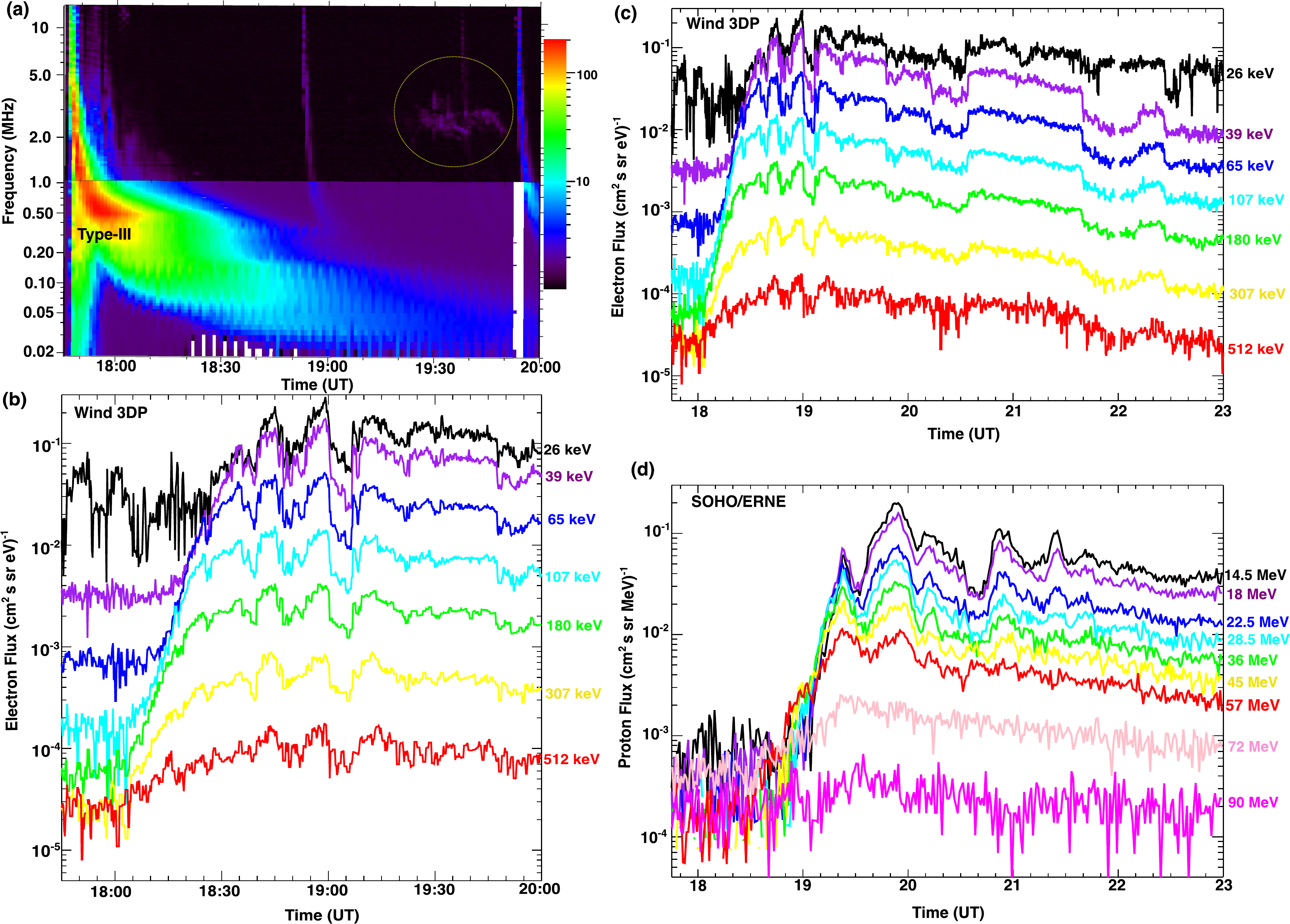}
}
\caption{{\bf Interplanetary radio bursts and solar energetic particles (electrons/protons) associated with Event E1}. (a) Wind/WAVES dynamic radio spectrum (1 min average) in the 0.02-13.82 MHz frequency range. The yellow circle highlights faint radio enhancement during CME propagation in the interplanetary medium. The color bar represents the average intensity of the electric field (in decibels (dB) above the background level). (b, c) Electron flux (12-s cadence) measured by Wind 3DP at different energies ranging from 26-512 keV. (d) Proton flux (1 min cadence) detected by SOHO/ERNE at different energies ranging from 14.5-90 MeV.} 
\label{fig2}
\end{figure*}

\section{Results}
\subsection{Event 1 (E1)}

\subsubsection{Plasma and magnetic configuration}
The active region (AR) NOAA 12017 was located at N10W32 and produced multiple C-class flares on March 29, 2014. New positive magnetic flux (P1) emerged in the preexisting AR on March 27 and formed a small fan-spine topology within a preexisting large pseudostreamer (Figure \ref{fig1}(a,b,c)). An enlarged view of the AR plasma configuration (AIA 211 \AA\ image at 15:05:35 UT on March 29, prior to an eruption) reveals a large null-point topology with a small, closed fan-spine structure embedded beneath the western lobe. The extrapolation shown in Figure \ref{fig1}(d) below was performed on an SDO/HMI magnetogram captured at 15:05:45 UT on March 29, 2014. The potential field extrapolation of the AR shows the pseudostreamer flux system (red field lines) and an embedded smaller fan-spine topology (cyan field lines) with outer and inner null points, N2 and N1, respectively (Figure \ref{fig1}(d)). An S-shaped filament existed (marked by an arrow in Figure \ref{fig1} (b)) within the inner fan-spine topology. A nonlinear force-free field extrapolation of the inner fan-spine topology shows sheared field lines at the filament site \citep{liu2015}. The inner and outer domes had approximate widths of 70$\arcsec$ and 330$\arcsec$, while the projected heights of the inner and outer nulls were around 50$\arcsec$ and 150$\arcsec$. The X-class flare associated with event E1 started at 17:35 UT, peaked at 17:48 UT, and ended at 17:54 UT.

\subsubsection{Pre-eruption activities}
During the pre-eruption phase, we observed multiple inflowing loop structures approaching the inner null (N1) (Figure \ref{fig1a} (a1-a4)). Simultaneously, recurrent brightenings were detected below the filament approximately two hours before the eruption (see Movie S1). A C1.1 flare was observed at the same polarity inversion line (PIL) at 16:25 UT, accompanied by a failed filament eruption (Figure \ref{fig1a}(a2,a3)). The filament rotated, interacted with the overlying strapping field, and remained confined as cool plasma drained back to the solar surface.

A side arcade (AIA 94~\AA) formed below N1 around 17:00 UT (see the time-distance (TD) intensity plot along R1S1, Figure \ref{fig1a}(c2,d1)) during magnetic reconnection at N1. The TD intensity plot along R2S2 shows  inflowing structures (1.4–2.0 \kms) moving toward the null (Figure \ref{fig1a}(b2,d2)). Simultaneous brightenings beneath the filament indicate heating, which led to the expansion of the overlying flux ($\approx$5 \kms) starting at $\approx$17:20 UT. These approaching flux systems encounter near the null (Figure \ref{fig1a}(d2)) and produced explosive energy release via reconnection at N1.
Note that the footpoints of these approaching structures are rooted in opposite polarities; therefore, they are oppositely directed flux systems, favorable for reconnection near the null.

\subsubsection{Signatures of breakout reconnection, flux rope formation/eruption, and associated fast EUV wave}
The AIA 304~{\AA} image at 17:40:31 UT (see Movie S2) shows footpoint brightenings (ribbons) which we interpret as signatures of ongoing reconnection between the approaching flux systems near the null point (Figure \ref{fig1b}(a1)). A cotemporal AIA 94~{\AA} image clearly reveals the formation of double side arcades (SA1 and SA2; Figure \ref{fig1b}(c1)). Simultaneous heating below the filament, observed as a hot sigmoidal feature, is also evident (Figure \ref{fig1b}(a1,b1)). AIA 131~{\AA} and 94~{\AA} images capture the formation of a hot circular structure, which we interpret as a flux rope (FR), along with a hot leading edge (LE) ahead of it (17:45–17:46 UT; Figure \ref{fig1b}(b2,b3,c2,c3)). Typical two-ribbon structures (J-shaped and inverse J-shaped, marked by R1 and R2) appeared below the erupting flux rope, along with a strong circular ribbon (R3) and a remote ribbon (R4) (Figure \ref{fig1b}(a2)).
Subsequently, the flux rope interacted with the ambient overlying field inside the pseudostreamer, leading to the formation of a bright, evolving side arcade (SA as indicated in Figure \ref{fig1b}(c4)) to the east of the typical flare arcade (FA). Additionally, a long hot plasma sheet (PS) extended above the SA (Figure \ref{fig1b}(b4,c4)), accompanied by recurrent upflows and downflows along the PS (Movie S2).

We analyzed RHESSI images to study the evolution of X-ray sources during the flare energy release. The RHESSI flux profiles reveal hard X-ray emission up to 300 keV during the impulsive phase (Figure \ref{fig1c}(c1)). During the pre-eruption phase, a 6–12 keV source was observed above the filament at 17:37:37 UT , coinciding with the heated overlying arcades observed in hot channels (Figure \ref{fig1b}(d1)). Another 6–12 keV source appeared around 17:43 UT, aligning with the side arcade SA1 (Figure \ref{fig1b}(d2)) during ongoing reconnection. During the impulsive phase associated with the eruption, a loop-top source (up to 12–25 keV) and two footpoint sources (25–50, 50–100, and 100–300 keV) were detected around 17:46:49 UT (Figure \ref{fig1b}(d3)). The footpoint sources coincided with flare ribbons R1 and R2 observed in AIA 304~{\AA} (Figure \ref{fig1b}(a3)).
During the post-eruption (flare decay) phase, a weak source at 30$\%$ of the peak intensity was observed at 18:07 UT in the 6–12 keV and 12–25 keV energy bands (Figure \ref{fig1b}(d4)).

The AIA 304~{\AA} images reveal the formation of a large-scale remote ribbon (R5) when the flux rope encountered the outer null (N2) at around 17:50 UT (Figure \ref{fig1c}(a1,a2)). To understand the evolution of the remote ribbons associated with reconnection at inner/outer nulls, we extracted peak counts from AIA 304~{\AA} images within the rectangular boxes around R4 and R5 in Figure \ref{fig1c}(a2). Intensity fluctuations were observed at R4 during the pre-eruption and eruption phases, with a peak around 17:46 UT as the flux rope approached the inner null N1 (Figure \ref{fig1c}(c2)). The intensity at R5 peaked around 17:50 UT.
AIA 211~{\AA} running-difference images and Movie S3 show a fast EUV wavefront ahead of the flux rope’s leading edge (LE) during its interaction near the inner null (N1) at approximately 17:47 UT (Figure \ref{fig1c}(b1)). Ribbon R5 is also detected in AIA 211~{\AA} running-difference images when the fast wave and flux rope propagate through N2 (Figure \ref{fig1c}(b2)). The flux-rope leading edge, which trailed behind the fast EUV wave, disappeared around 17:48 UT (see Movie S3). New slow EUV wavefronts appeared after the fast EUV wave passed through N2 (Figure \ref{fig1c}(b3,b4)).
The TD intensity plot along R3S3 shows the flux rope (FR) and leading edge (LE) moving at speeds of about 500 \kms and 680 \kms, respectively (Figure \ref{fig1c}(d1)). The TD intensity plot along R4S4 reveals fast and slow EUV wavefronts traveling at speeds $\approx$1008 \kms and $\approx$428 \kms, respectively (Figure \ref{fig1c}(d2)).
Observations at 410 MHz from the Sagamore Hill Radio Observatory display quasiperiodic decimetric Type III bursts (Figure \ref{fig1c}(c3)) during the flux rope interaction near the inner and outer nulls (17:45–17:53 UT).

\subsubsection{Radio bursts and CME}

The flux-rope eruption produced a halo CME, which appeared in the LASCO C2 field of view at 18:12 UT with an initial speed (plane of sky) of $\approx$528 \kms. A strong shock front was observed ahead of the CME (Figure \ref{app-fig1a}). The CME decelerated at a rate of 4.13 m s$^{-2}$ in the LASCO C2/C3 field of view. We observed a slowly moving blob in LASCO C2 coronagraph images prior to the CME. The CME shock interacted with the blob moving ahead of it at about 6 R$_{\odot}$ (Figure \ref{app-fig1a}(e,f)).

The dynamic radio spectrum from e-Callisto (Roswell Observatory) in the 45–81 MHz and 220–450 MHz ranges reveals a series of Type III bursts occurring between 17:45:30 and 17:58 UT (Figure \ref{app-fig1b}(a,b)). The 1-second cadence flux-density profile at 410 MHz from RSTN shows quasiperiodic Type III bursts, consistent with the e-Callisto spectrum. Additionally, a metric Type II burst was detected at 240–300 MHz (harmonic) around 17:47 UT. The lower-frequency spectrum (45–81 MHz) also exhibits a Type II burst (fundamental), which is more clearly observed in the radio dynamic spectrum from the Sagamore Hill station (25-180 MHz) between 17:47 and 18:00 UT.
 Using the Newkirk density (one-fold) model \citep{newkirk1961}, the estimated shock speed derived from the drift rate of the Type II burst (fundamental emission) was 1488 \kms and 1194 \kms for two different lanes (marked as 1 and 2, Figure \ref{app-fig1b}(c)). These values are higher than the projected shock speed of 1008 \kms measured in EUV images, which include projection effects unlike the radio emission. 

\subsubsection{Interplanetary radio bursts and SEPs}
SEPs (electrons/protons) observations are displayed in Figure~\ref{app-fig1aa}. For the event E1 analyzed in this section, a large SEP event with onset at $\sim$18:00~UT followed by a long decay was observed.
The Wind/WAVES dynamic radio spectrum (0.02–13.8 MHz) reveals multiple Type III bursts (Figure \ref{fig2}(a)) during the eruption (17:48–18:00 UT). A radio enhancement, characterized by a Type II burst, was observed in the 2–3 MHz band from 19:25–19:50 UT, as the CME shock interacted with a pre-existing slowly moving blob at about 7 R$_\odot$ in the interplanetary medium.

The Wind 3DP instrument detected energetic electrons (26-512 keV) at 1 AU, with high-energy electrons at 512 keV showing onset around 18:04 UT (Figure \ref{fig2}(b,c)). Clear quasiperiodic behavior with at least five peaks was observed in the flux profile between 18:00 and 19:30 UT, and the electron enhancement continued until about 23:00 UT. The SOHO ERNE instrument observed energetic protons (14.5-90 MeV) at 1 AU, with high-energy proton flux at 90 MeV starting to rise around 18:50 UT (Figure \ref{fig2}(d)). Quasiperiodic behavior in proton fluxes (14.5-57 MeV) was evident from 18:50 to 20:50 UT.

Velocity Dispersion Analysis (VDA) is a commonly used method \citep{krucker2000,reames1999,vainio2013} to estimate the solar particle release (SPR) time at the Sun by assuming that particles of different energies are simultaneously released at the Sun and travel scatter free along the same magnetic field line and arrive at 1 AU with energy-dependent velocities. The arrival time $t(E)$ of a particle with energy $E$ is given by: $t(E) = t_0 + \frac{L}{v(E)}$, where $t_0$ is the release time at the Sun, $L$ is the path length traveled by the particles, and $v(E)$ is the velocity of particles with energy $E$. By plotting the observed onset times of particles against their inverse velocities ($1/v$), the solar release time $t_0$ can be estimated from the intercept of a linear fit. The slope represents the path length along the Parker spiral. The observed onset time at 1 AU is considered to correspond to the first-arriving particles for each energy. Any deviation from these assumptions, such as scattering, extended release times, or variable path lengths, can introduce uncertainties in the estimated SPR time.

We utilized Wind 3DP electrons (26-307 keV) to determine the SPR time and path length using VDA analysis. We estimated the velocity ratio \(\beta = v/c\) for near-relativistic electrons and protons using the relativistic energy-momentum.
Figure \ref{app-fig6} displays a plot of $\beta^{-1}$ versus onset time for electrons. The estimated SPR time is 17:48 UT$\pm$2.6 min, with a path length of 1.53$\pm$0.14 AU. The SPR time is consistent with the onset of quasiperiodic decimetric/metric Type III bursts, shock formation, and the associated Type II burst during inner breakout reconnection.

SOHO orientation during this period was not appropriate for the particle instruments on board this spacecraft to detect the first arriving particles if these propagated along nominal Parker spiral magnetic field lines (the nominal roll angle of SOHO was switched to 180$^{\circ}$ on 2014 March 24, five days before the SEP event).
A velocity dispersion analysis did not provide reliable proton release times. Therefore, we utilized Time Shift Analysis (TSA), assuming an average path length of $L\approx1.2~AU$. For 72 MeV protons ($v=0.37c$), the travel time is given by $t = \frac{L}{v}\approx 27~min$. Subtracting this from 1 AU (18:45 UT $\pm$3 min) onset time suggests a solar release at approximately 18:18 UT$\pm$3 min. This proton release time is inconsistent with the electron release timing. There is a possibility that protons might release earlier, i.e., simultaneously with electrons. However, given the uncertainties in interplanetary transport conditions (scattering, pitch angle distribution) and the inadequate SOHO/ERNE field of view, our current analysis cannot rule out this possibility. In principle, the faster electrons experience less scattering processes than the slower protons. In addition, electrons have the smallest pitch angle, as measured by Wind/3DP, indicating that they propagated along or near the observed magnetic field lines.

\begin{figure*}
\centering{
\includegraphics[width=18cm]{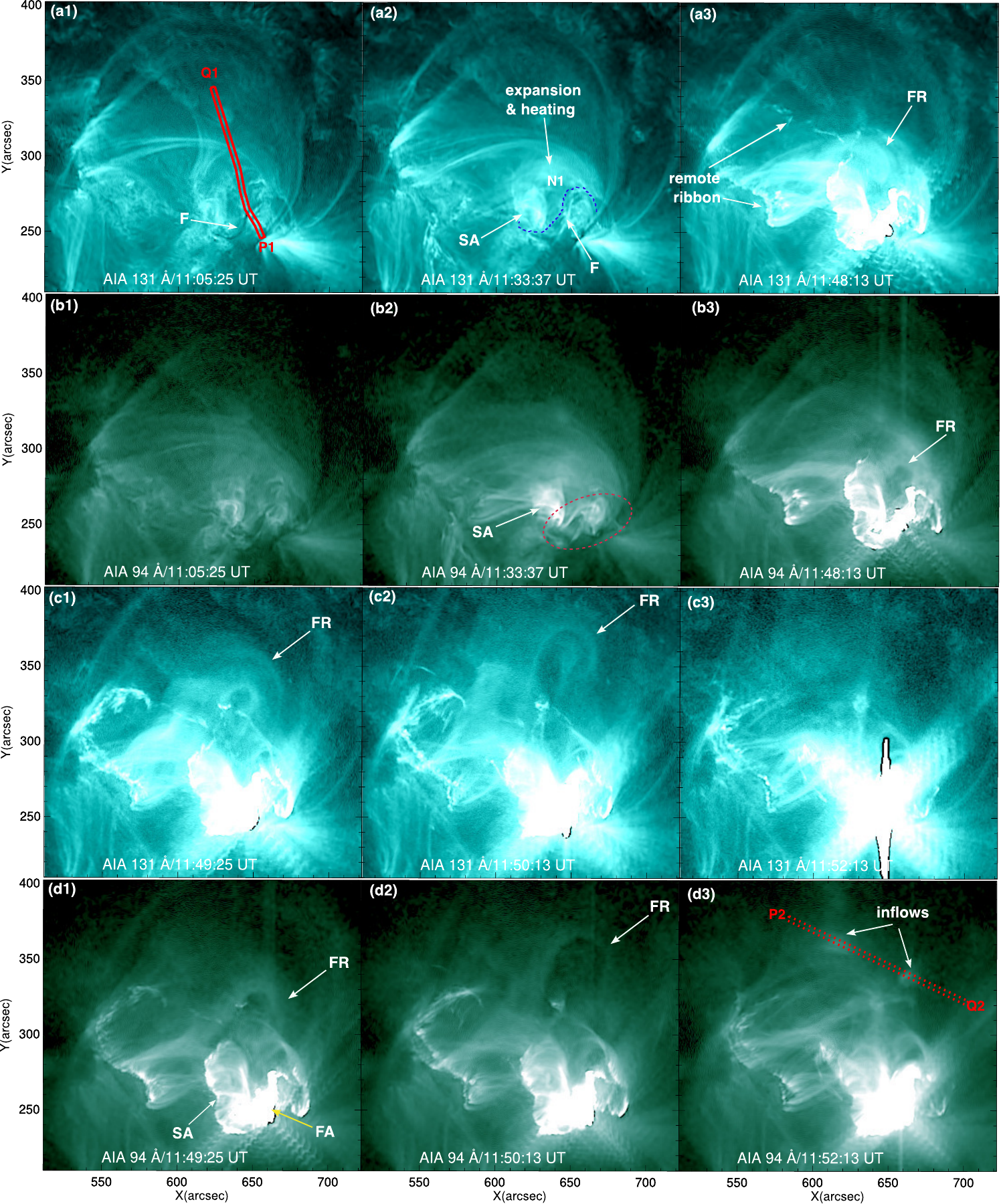}
}
\caption{{\bf Formation and eruption of a flux rope along with breakout/interchange reconnection during Event E2 on March 30, 2014}. AIA 131 and 94~{\AA} images depict the flux rope (FR) formation and eruption. `F' denotes the filament. `N' marks the approximate position of the inner null. `SA' indicates the side arcade, and `FA' represents the flare arcade. The red ellipse in panel (b2) highlights the sigmoidal structure. The slits P1Q1 and P2Q2 are used to create the time-distance intensity maps in Figure \ref{ribb-aia}(c1, c2, c3). }
\label{fig-aia}
\end{figure*}
\begin{figure*}
\centering{
\includegraphics[width=18cm]{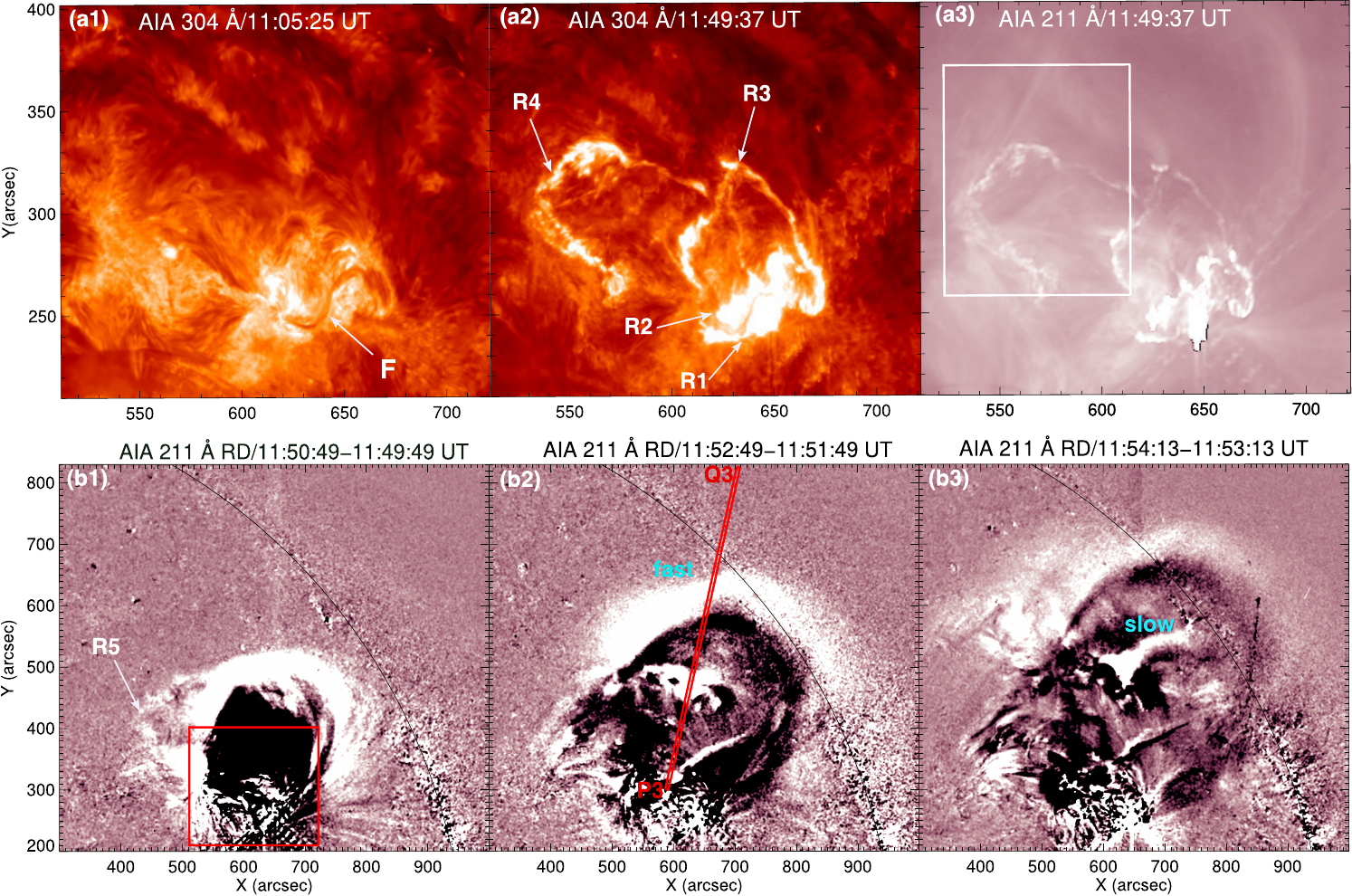}
\includegraphics[width=18cm]{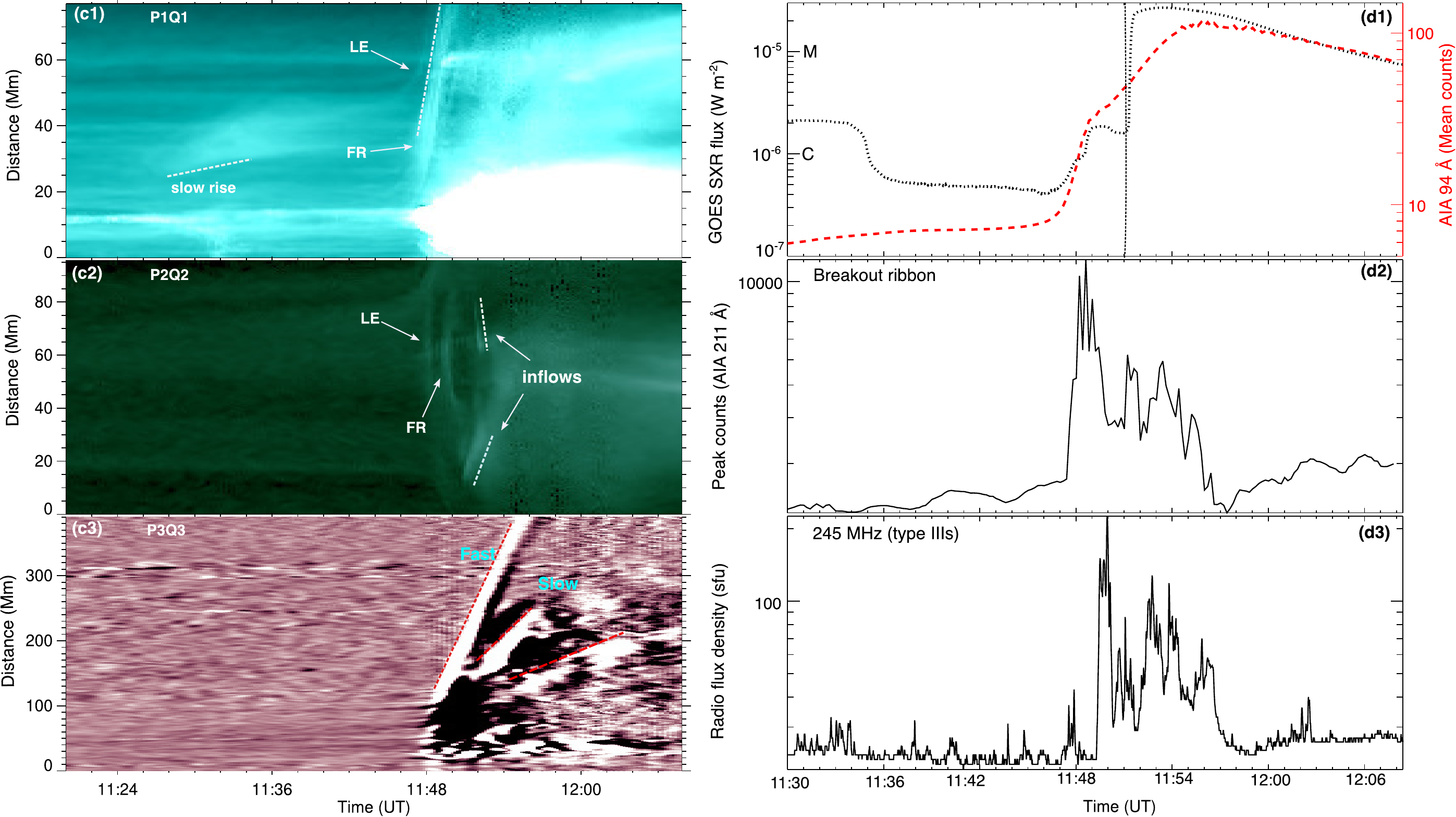}
}
\caption{{\bf Kinematics of the flux rope and associated fast/slow EUV wavefronts during Event E2}. 
(a1–a3) AIA 304 and 211~{\AA} images showing the filament and multiple ribbons.
(b1–b3) AIA 211~{\AA} running-difference images illustrating the fast and slow EUV waves. The red box indicates the field of view for the top-row panels. 
(c1–c3) Time-distance intensity plots along slits P1Q1, P2Q2, and P3Q3 (LE=leading edge, FR=flux rope).
(d1–d3) GOES soft X-ray flux (1–8~{\AA}) and AIA 94~{\AA} peak counts (red dashed curve) extracted from the eruption site (red box in (b1)). The vertical dashed line indicates the onset of energy release at N2. The AIA 211~{\AA} peak counts in (d2) are extracted from the breakout ribbons (R3, R4) within the white box shown in (a3). RSTN radio flux density (1-second cadence) at 245 MHz showing quasiperiodic Type III radio bursts.} 
\label{ribb-aia}
\end{figure*}

\subsection{Event 2 (E2)}
\subsubsection{Pre-eruption activities}
According to GOES soft X-ray flux (1–8~{\AA}) measurements, an M2.1 flare began at 11:48 UT, peaked at 11:55 UT, and ended around 12:02 UT on March 30, 2014 (Figure \ref{ribb-aia}(d1)).
We observed the simultaneous activation of the filament, a gradual expansion of the overlying flux near the inner null (N1), and associated heating signatures at around 11:27 UT (Figure \ref{fig-aia}(a1,a2,b1,b2)). A side arcade (SA) east of the filament brightened at the same time, well before the formation of the flux rope and onset of the eruption. Moreover, we detected weak Type III radio bursts during the pre-eruption phase.

\subsubsection{Signatures of breakout reconnection, flux rope formation/eruption, and associated fast EUV wave}
The AIA 131 and 94~{\AA} images reveal the formation and ejection of a flux rope during the eruption. A small circular feature, which we interpret as a flux rope (FR), appeared above the inner system's PIL, where an S-shaped filament and a sigmoidal hot structure were located (Figure \ref{fig-aia}(a2,a3,b3)). Circular and remote ribbons were observed as the flux rope interacted with the overlying pseudostreamer field near the inner null. Following this interaction, the flux rope's bubble-like structure expanded as it transited the PS (Figure \ref{fig-aia}(c1,c2,d1,d2)). Inflows were detected below the flux rope as it accelerated outward (Figure \ref{fig-aia}(d3)). The flare arcade (FA) beneath the flux rope and the side arcade (SA) expanded as the flux rope erupted. The flux rope was visible only in the hot channels (131~ \AA~ and 94 \AA), and no filament material was detected within the flux rope during its formation and eruption. AIA cool (304 \AA) and warm (211 \AA) channels show brightening within the filament during the pre-eruption activation but do not show any evidence of escaping filament material (see Movie S4). Therefore the bulk of the filament remained below the flux rope throughout the eruption.

The formation and eruption of the flux rope produced typical two-ribbon structures (R1, R2), presumably through reconnection in a flare current sheet beneath the flux rope (Figure \ref{ribb-aia}(a1-a3)). When the flux rope encountered the overlying strapping field near the inner null (N1), a quasi-circular ribbon (R3) and a remote ribbon (R4) simultaneously appeared (11:48-11:49 UT, Figure \ref{ribb-aia}(a2)), interpreted as breakout ribbons at the fan-spine separatrix and outer spine footpoints, respectively. Subsequently, as the flux rope interacted with the outer null (N2), another faint remote brightening (R5) was observed (11:50:49 UT, Figure \ref{ribb-aia}(b1)). AIA 211~\AA\ running-difference images, along with the accompanying animation (Movie S5), reveal a fast EUV wavefront ahead of the flux rope during magnetic reconnection at the inner null (N1). Two slow EUV wavefronts trailed the fast EUV wave, with one of the slow fronts disappearing around 11:56 UT (Figure \ref{ribb-aia}(b1-b3)).

The TD intensity plot along the slit P1Q1 indicated in Figure \ref{fig-aia}(a1) (AIA 131~\AA) shows pre-eruption activity near N1 between 11:26 and 11:36 UT (Figure \ref{ribb-aia}(c1)), including the expansion of the loops above the filament and associated heating. The curved path of slice P1Q1 was selected to track the nonradial motion of the flux rope within the inner fan-spine topology. The estimated speed of the expanding loops, which became the leading edge (LE) of the flux rope later, was about 13~\kms. Simultaneously, brightening was observed within the filament channel. The flux rope (FR) rose at about 470~\kms during reconnection through N1.
Multiple faint outflows were observed along the plasma sheet during 11:49 and 11:58 UT, following the flux rope's escape into the outer corona (Figure \ref{ribb-aia}(c1)). Strong inflows were detected below the flux rope in the AIA 94~{\AA} channel (11:51–11:53 UT) shortly after the ejection. The estimated inflow speeds from the TD intensity plot along slit P2Q2 (indicated in Figure \ref{fig-aia}(d3)) were $\approx$326 \kms and $\approx$240 \kms (Figure \ref{ribb-aia}(c2)).
The fast EUV wave in the TD running-difference intensity plot along slit P3Q3 (indicated in Figure \ref{ribb-aia}(b2)) propagated ahead of the erupting flux rope at a speed of $\approx$744 \kms (Movie S5). The two slow EUV fronts propagated at projected speeds of 328 \kms and 178 \kms, respectively (Figure \ref{ribb-aia}(c3)).

The mean AIA 94~{\AA} intensity extracted from the flaring site (red dashed curve in Figure \ref{ribb-aia}(d1)) is temporally correlated with the M-class flare observed in the GOES soft X-ray flux (1–8~\AA). Ribbon peak intensity was determined by summing the pixel intensities within a  box that covers the remote ribbons (R3, R4) in the AIA 211~{\AA} images (Figure \ref{ribb-aia}(a3, d2)). During the flare's impulsive and main phases, most AIA channels were saturated, with AIA 211~{\AA} being the least affected. Therefore, this channel was utilized to estimate the ribbon intensity, which is interpreted as a proxy for the recurrent precipitation of downward-moving electrons during magnetic reconnection at N1.
The intensity profile reveals a clear quasiperiodic decaying oscillation with five peaks (11:47–11:57 UT). Interestingly, the perturbations in circular and remote-ribbon intensity match the Type III radio bursts detected at 245 MHz by RSTN (Figure \ref{ribb-aia}(d3)).

\subsubsection{Radio bursts and CME}
\begin{figure*}
\centering{
\includegraphics[width=18cm]{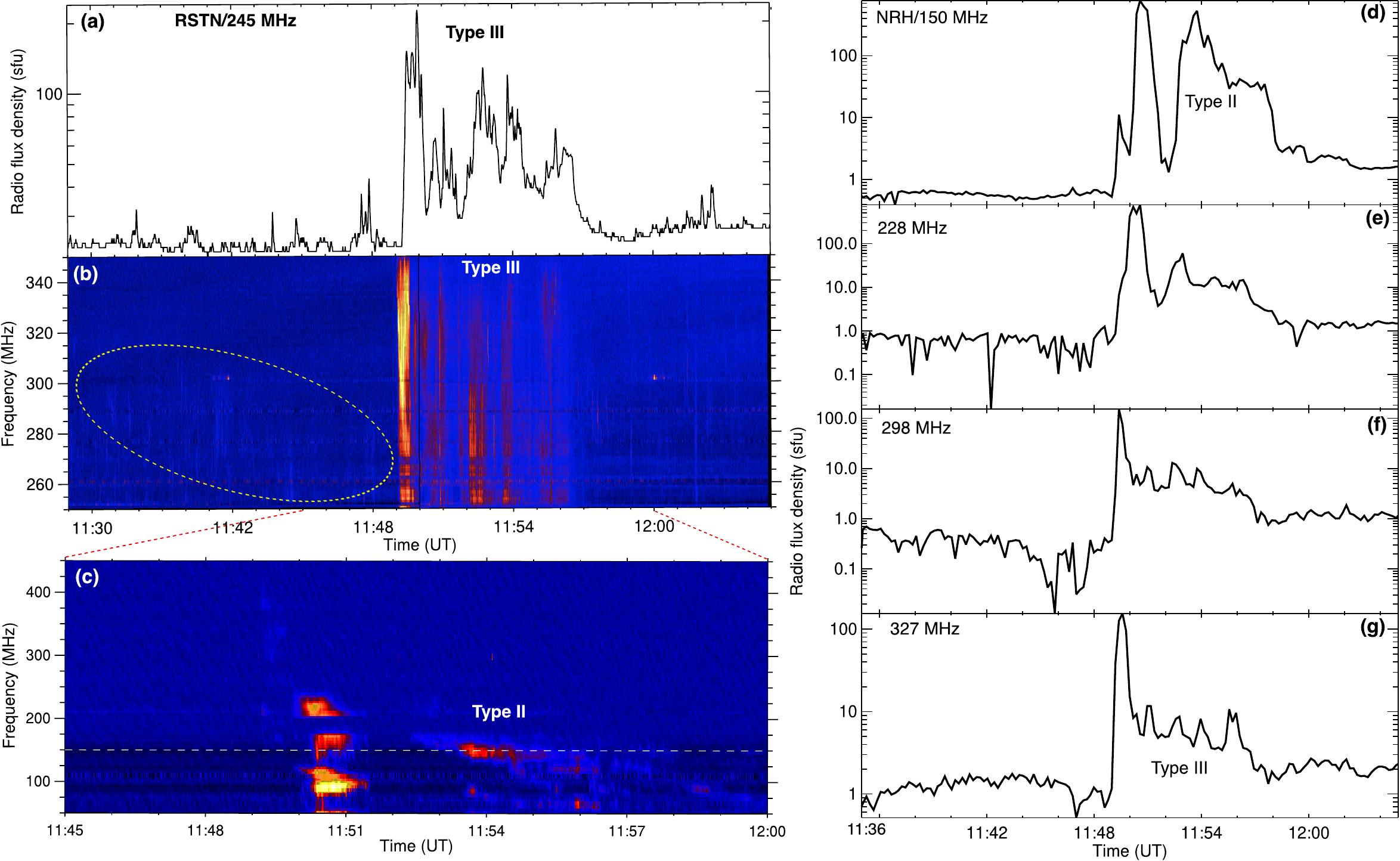}
}
\caption{{\bf Radio bursts before and during Event E2.} 
(a) RSTN radio flux density (sfu, 1-s cadence) at 245 MHz.
(b, c) E-Callisto dynamic radio spectra from KRIM (Crimean Astrophysical Observatory) and MRT1 (Mongolia) stations. The oval in (b) highlights the faint radio emission during the pre-eruption phase. The horizontal dashed line in (c) indicates the 150 MHz frequency.
(d, e, f, g) Radio flux density (12-s cadence, sfu) extracted from NRH images of the northwest quarter of the Sun at four frequencies between 150 and 327 MHz during 11:45-12:00 UT.} 
\label{fig3}
\end{figure*}

The radio observations of Event E2, including the dynamic radio spectrum and RSTN flux measurements, reveal a series of faint Type III bursts at 245–300 MHz during the pre-eruption phase (11:30–11:48 UT; Figure \ref{fig3}(a,b)). During the impulsive phase, a series of quasiperiodic, strong Type III bursts at 250–350 MHz were observed between 11:49 and 11:57 UT. The dynamic radio spectrum also contained a metric Type II (second harmonic) radio burst at 50–175 MHz during 11:51–12:01 UT (Figure \ref{fig3}(c)).

We utilized radio imaging observations at different frequencies (150–327 MHz) from the NRH to extract the radio flux density within the eruption site  (northwestern field of view), integrating the images over 12 seconds to match the AIA cadence. The radio flux profiles  exhibit Type III and Type II radio bursts at 150 MHz (Figure \ref{fig3}(d)). The flux profiles at higher frequencies (228–327 MHz) also exhibit quasiperiodic Type III bursts cotemporal with the Type II, consistent with those observed in the e-Callisto dynamic radio spectrum (Figure \ref{fig3}(e,f,g)).

\begin{figure*}
\centering{
\includegraphics[width=18cm]{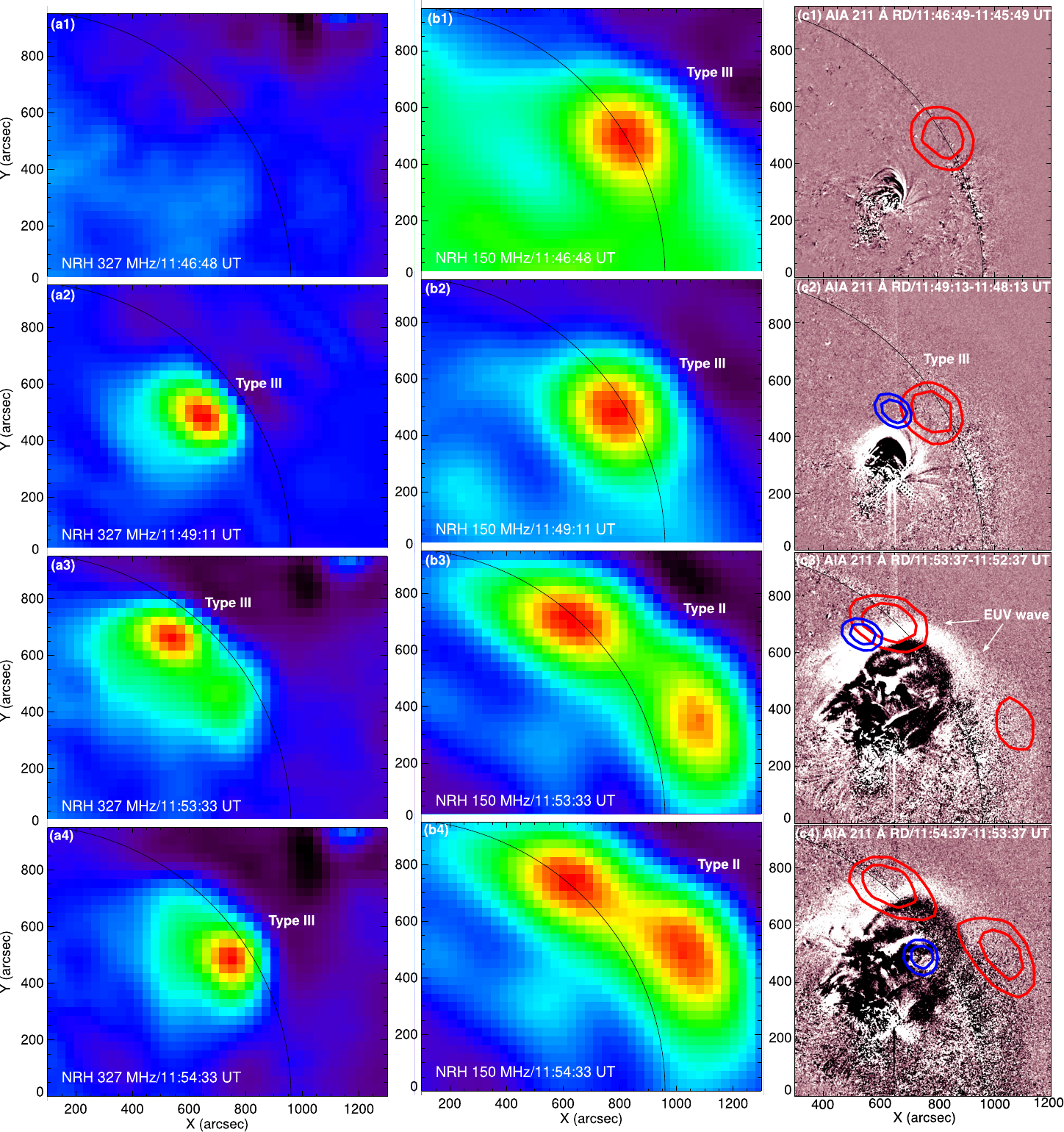}
}
\caption{{\bf Imaging of radio sources during Event E2.} 
(a1-a4) NRH radio sources (Type III bursts) at 327 MHz.
(b1-b4) NRH radio sources at 150 MHz (Type III and Type II bursts).
(c1-c4) Contours of radio sources at 150 MHz (red) and 327 MHz (blue) overlaid on AIA 211 {\AA} running difference images ($\Delta$t = 1 min). The contour levels are 80\% and 90\% of the peak intensity. The black curve traces the solar limb.} 
\label{fig4}
\end{figure*}

Figure \ref{fig4} and the associated animation (Movie S6) illustrate the temporal evolution of the imaged NRH radio sources during 11:30–12:03 UT, with radio contours at 150 MHz and 327 MHz overlaid on AIA 211~\AA\ images. These observations cover both the pre-eruption and the impulsive phases. At 150 MHz, the images reveal a pre-eruption Type III burst at $\approx$11:46:48 UT (Figure \ref{fig3}(d) and \ref{fig4}(b1)).
However, faint Type III-like features observed in the 245–300 MHz range in the dynamic radio spectrum were not detected in the NRH images (Figure \ref{fig3}(d–g)), likely due to the limited resolution of NRH. Stronger Type III sources appeared during the flux rope's interaction with the ambient magnetic field, which are interpreted by explosive magnetic reconnection near the inner null point (11:49 UT, Figure \ref{fig4}(a2, b2, c2)).
The Type II radio burst (double sources at 150 MHz in the bottom rows of Figure \ref{fig4}) was co-spatial with the fast EUV wave propagating ahead of the erupting flux rope. The animation (Movie S6) further shows the lateral expansion of the radio sources alongside the EUV wave, particularly the northern component, during the Type II burst (11:53–11:55 UT).

The CME was observed by the LASCO C2 (2–6 R$_\odot$) and STEREO-B COR1 (1.4–4 R$_\odot$) coronagraphs. The running-difference images  clearly show the lateral expansion of faint wavefronts to the north and south (indicated by arrows in Figure \ref{app-fig2}(a1-a3,b1-b3)). The flux rope structure is prominently visible and aligns well with the flux rope observed in the AIA 131~{\AA} channel (Figure \ref{app-fig2}(a3,b3)). The CME had a speed (plane-of-sky) of 486 \kms and an acceleration of -3.02 m s$^{-2}$ within the LASCO C2/C3 field of view.
Metric Type III and Type II (fundamental) radio bursts were detected in the 25–180 MHz range by the RSTN San Vito station between 11:50 and 12:04 UT (Figure \ref{app-fig2}(c1)). Interplanetary radio data from Wind/WAVES (0.02–13.8 MHz) contained a series of quasiperiodic Type III bursts during the same period (Figure \ref{app-fig2}(c2)).

\subsection{Event 3 (E3)}
This eruption was a hybrid jet-CME event, which began as a jet in the low corona associated with a filament eruption from the east side of the circular PIL of the inner fan-spine topology. The M2.0 flare started at 19:04 UT, peaked at 19:18 UT, and ended at 19:27 UT on March 28) (Figure \ref{app-fig4}(c1)). 
During the pre-eruption phase, a confined filament eruption was observed on the western side of the circular PIL within the closed fan-spine topology (17:35–17:40 UT). The filament kinked and was halted by the overlying strapping field, preventing it from reaching the inner null (Figure \ref{app-fig3}(b1-b3)).  The overlying pseudostreamer flux opened up and produced coronal dimming above the inner fan-spine topology (18:02–18:14 UT) (Figure \ref{app-fig3}(a1-a3)). This activity was followed by signatures of slow magnetic reconnection at the inner null, accompanied by ribbon brightening and the gradual rise of the filament at the eastern PIL (18:15–19:00 UT). Brightening below the filament accompanied the formation of a small flux rope around the filament, as observed in the AIA 94~{\AA} channel (see Movie S7).

The first confined jet was observed along the closed outer spine during the pre-eruption phase during slow reconnection at the inner null between 19:01 and 19:08 UT (Figure \ref{app-fig3}(b4); also see 94~\AA\ images in Movie S7). Later, an explosive jet was initiated when the flux rope encountered the inner null at approximately 19:11 UT (Movie S7), accompanied by the formation of flare ribbons (R1, R2), a bright circular ribbon (R3), and a remote ribbon (R4) at the end of the outer spine (Figure \ref{app-fig4}(c1)). Metric Type III radio bursts were detected at 19:09 UT and between 19:12 and 19:23 UT during the jet and subsequent flux-rope eruption (Figure \ref{app-fig4}(c3)). The side arcade above the western PIL also reconnected through the inner null, producing long-duration recurrent jets and plasma blobs during the post-eruption phase (19:20-19:50 UT). These jets are best observed in the AIA 304~\AA\ and 211~\AA\ channels (Movie S7).

AIA 211~{\AA} running-difference images and the accompanying animation (Movie S8) reveal the formation of a large-scale EUV wavefront shortly after the confined jet ($\approx$19:14 UT onwards), associated with explosive reconnection at the inner null (Figure \ref{app-fig4}(a1-a3)). The estimated projected speed of the EUV wave, which we interpret as an MHD shock, was about 640 \kms, consistent with the observed metric Type II burst associated with the wave (Figure \ref{app-fig4}(c4)). We observed remote ribbon R4 at the eastern footpoint of the outer spine and another large-scale ribbon, R5, when the shock and erupting flux rope encountered the outer null (Figure \ref{app-fig4}(b1)).
Recurrent metric Type III bursts (245 MHz) were detected between 19:09 and 19:22 UT (Figure \ref{app-fig4}(c3)). These bursts aligned with the EUV intensity extracted from ribbon R3, serving as a proxy for recurrent magnetic reconnection (Figure \ref{app-fig4}(c2)). The Type III bursts were observed during interchange reconnection at both the inner and outer nulls.

The eruption produced a slow CME, which first appeared in the LASCO C2 field of view at 20:00 UT. The CME propagated with a projected speed of 420 \kms and an acceleration of 2.4 m s$^{-2}$. LASCO C2 coronagraph images reveal a faint shock associated with the CME (Figure \ref{app-fig4}(b2,b3)), which is also clearly visible in the STEREO-A COR2 field of view (2–15 R$_{\odot}$) (Figure \ref{app-fig4}(b4,b5)).
The dynamic spectrum from the Sagamore Hill Radio Observatory shows narrow-band Type III bursts (19:10–19:45 UT) and a metric Type II radio burst (19:20–19:35 UT) associated with the eruption (Figure \ref{app-fig4}(c4)). The estimated shock speed, derived from the drift rate of the Type II burst, was 528 \kms.

\section{DISCUSSION}\label{discussion}
We analyzed a series of eruptions from a nested-null topology from March 28-30, 2014. These topologies form through flux emergence beneath one side of a pseudostreamer, which appeared in this case to be a single-null fan-spine configuration. The newly emerged small embedded bipole, one day before (March 27), formed a closed fan-spine structure within one lobe of the pseudostreamer. The recurrent eruption of the inner fan-spine topology produced multiple jets, flares, CMEs, and shocks. Eruptions E1 and E2 originated from the same section of the PIL, specifically from an S-shaped filament channel. Event E4 (see Table 1 and the animation below) is similar to Event E3 and is therefore not discussed in the previous section. Both eruptions originated from the same PIL and produced jets accompanied by multiple blobs \footnote{\url{https://hesperia.gsfc.nasa.gov/sdo/aia/movies/2014/03/28/20140328_2345-0016/}}, although the associated CME and shock were slightly faster than those in E3. Confined eruptions were also observed prior to E1 and E3 from the western section of the PIL, followed by successful eruptions. 

\subsection{Pre-eruption signatures and formation of flux ropes}
The observations of the event E1 reveal inflow structures moving toward the inner null, N1, at a speed of $\approx$1.4–2.0 \kms. Simultaneously, we observed brightening below the filament and the expansion of the overlying field ($\approx$5 \kms), which reached the inner null before the onset of explosive breakout reconnection. Additionally, a side arcade formed during slow breakout reconnection at the inner null, about 40 minutes before the filament eruption.

In Event E2, weak Type III radio bursts, interpreted as signatures of escaping electron beams, were detected during the pre-eruption phase ($\approx$15 minutes before the filament eruption) and were linked to slow breakout reconnection at N1.

Event E3 was characterized by pre-eruption opening and dimming near the outer null, N2, followed by slow breakout reconnection at N1, about one hour before the filament eruption. These observations suggest that the overlying field was gradually removed via slow breakout reconnection before the filament eruption. These pre-eruption activities are consistent with signatures (pre-eruption jets, dimmings) observed in previous eruptions in pseudostreamers and fan-spine topologies \citep{kumar2019b,kumar2021,kumar2024a,karpen2024} as well as in MHD simulations of single null-point topologies \citep{lynch2013,wyper2024}.

In all events, flare reconnection below the filament contributed to the flux rope's formation during the filament eruption. The flux rope was only detected in the AIA hot channels (AIA 131/94~{\AA} channels), and flare reconnection heats the rope up to about 10 MK. Flare reconnection also created two ribbons (R1, R2) below the erupting flux rope.

\subsection{Breakout reconnection}
These observations suggest that all events follow a similar eruption process and exhibit consistent multiwavelength signatures (e.g, ribbon formation and associated radio bursts). Characteristics of all eruptions are summarized in Table 1. In all homologous events, a quasicircular ribbon (R3) appeared at the footpoints of the fan separatrix and a remote ribbon (R4) appeared at the footpoint of the outer spine during explosive breakout reconnection at the inner null, N1. Because the internal pseudostreamer magnetic field and the flux rope were both closed, this breakout interaction involved closed-closed reconnection and not interchange. The flux rope accelerated as the overlying strapping field was removed, and a low coronal shock appeared ahead of the flux rope. Another large-scale remote ribbon (R5) appeared via a second phase of breakout reconnection when the flux rope and shock encountered the outer null, N2. Note that this breakout reconnection was also interchange, because the pseudostreamer was surrounded by open flux.
The intensity at remote ribbons R4 and R5 peaked during explosive breakout reconnection at the inner and outer nulls, respectively. For example, in event E1, the peak intensity at R4 preceded the peak at R5 by about 4 minutes (Figure \ref{fig1c}(c2)). The estimated distance between R4 and R5 is about 240$\arcsec$, yielding a source speed of $\approx$750 \kms. This speed is comparable to the typical velocity of the flux rope's leading edge in the low corona for event E1.

Quasiperiodic decaying Type III radio bursts occurred during the interaction of the flux rope with the overlying magnetic flux, corresponding to breakout reconnection through the breakout current sheet (BCS). At the same time, a quasiperiodic enhancement in the flux of the breakout ribbon (circular/remote) suggests recurrent precipitation by downward-moving electron beams. These observations indicate the simultaneous injection of bidirectional electron beams, both sunward and anti-sunward. The BCS transitioned to a flare current sheet (FCS) as the flux rope is ejected outward into the interplanetary medium \citep[e.g., ][]{lynch2013,wyper2024}, and multiple outflows were observed during and after the breakout reconnection.

In event E3, the flux rope was partially destroyed via breakout reconnection at the inner null and produced a jet. The remaining flux rope transitioned into a CME with a leading edge and a shock front ahead of it. The flux rope (circular feature) was detected in the COR2 images. This event serves as a good example of a hybrid jet-associated shock (and Type II bursts) formed via breakout reconnection at the inner null. Similar cases have been reported in previous studies involving fan-spine topologies (even without CMEs) \citep{kumar2013,kumar2015,kumar2016}. A series of recurrent jets without CMEs followed when the overlying field and filament channel from the opposite side of the first eruption reconnected through N1.

Most pseudostreamers produce narrow jets and slow CMEs \citep{wang2015,wang2023}. However, a few case studies \citep{kumar2021} have reported fast and wide CMEs ($\approx$1000 \kms) accompanied by shocks (in EUV and white-light images) originating from pseudostreamers (i.e., single null-point topologies). In the present study, we observed radio-loud, moderate-speed CMEs and associated shocks from nested-null topologies that were detectable from the low corona out to the interplanetary medium.

\subsection{Quasiperiodic Pulsations}
Quasiperiodic pulsations (QPPs) are commonly observed in solar and stellar flares, often linked to mechanisms such as recurrent magnetic reconnection associated with plasmoids or MHD oscillations \citep{nakariakov2009,Zimovets2021}. In null-point topologies, QPPs can result from periodic reconnection near magnetic nulls, where oscillatory energy release leads to modulated particle acceleration along with fast-mode waves \citep{mclaughlin2018,kumar2016a,kumar2017}. 

In all 4 events, QPPs were found in radio and EUV channels during the interaction of the erupting flux ropes with the overlying magnetic field through breakout reconnection near both inner and outer nulls. A wavelet analysis \citep{torrence1998} of the radio flux densities in the Type III bursts for event E1 reveals periodicities of 40 and 100 seconds (Figure \ref{app-fig5}). Event E2 exhibits a 100-second periodicity, while event E3 shows periodicities of 25 and 100 seconds. The 100-second period is present in all events, suggesting a common characteristic of breakout reconnection in this active region.    

In typical null-point topologies, the rising flux rope stretches and distorts the overlying magnetic field, leading to repetitive episodes of magnetic reconnection near the distorted null along with plasmoids \citep{kumar2019b,wyper2022}. This reconnection can inject energy into the surrounding plasma, generating bursts of accelerated particles and heating.  QPPs in the observed radio and EUV emissions indicate that the reconnection is inherently bursty and quasiperiodic, which may be modulated by plasmoids or fast MHD waves \citep{nakariakov2009,kumar2025}. The outward-moving electrons produce quasiperiodic radio bursts, while the downward-moving electrons generate quasiperiodic emissions at the circular and remote ribbons. The existence of two breakout current sheets in such nested-null configurations adds more opportunities for particle acceleration, remote footpoint heating, and particle escape into the heliosphere through reconnection at the outer BCS.

\begin{figure*}
\centering{
\includegraphics[width=18cm]{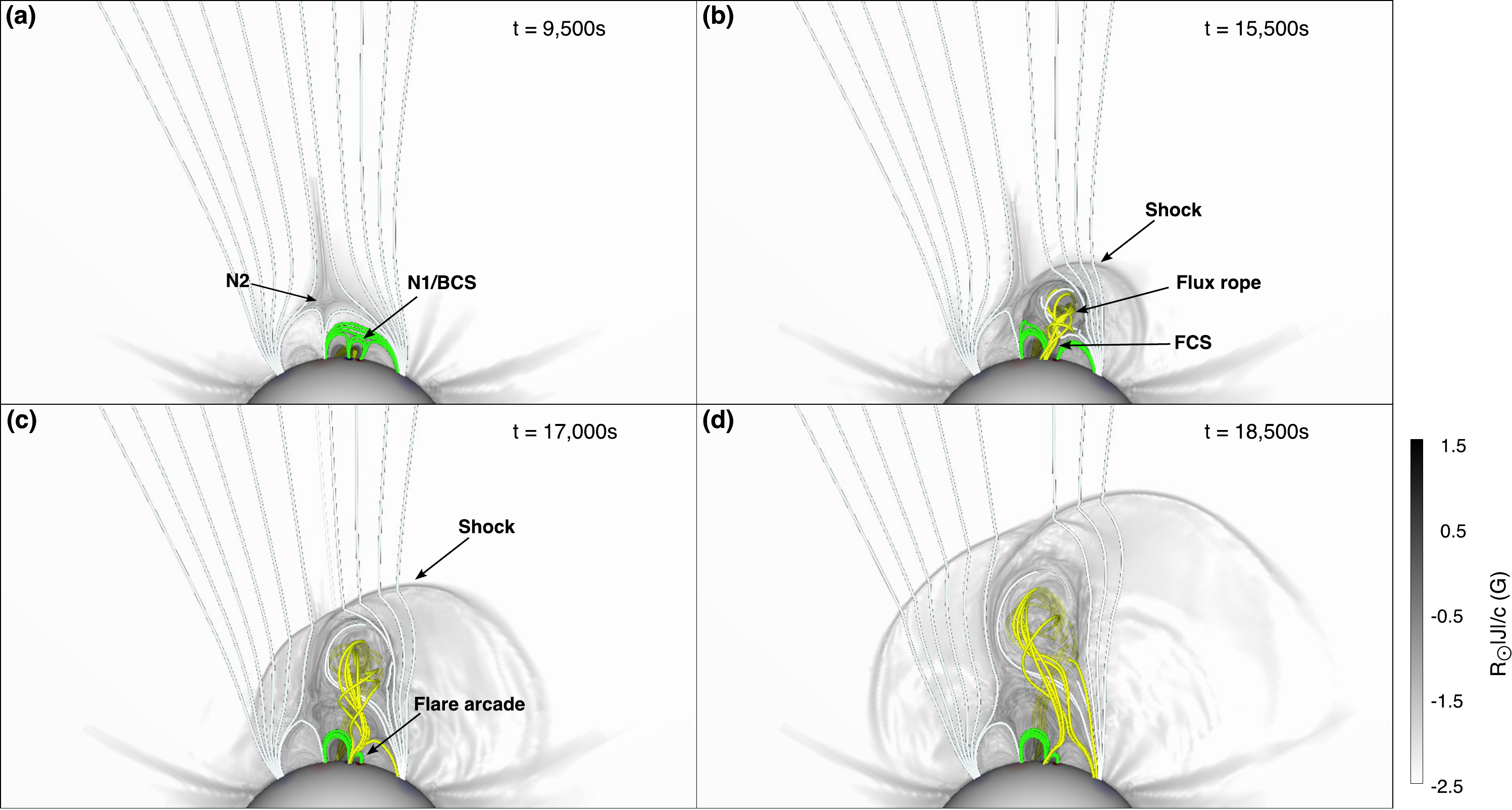}
}
\caption{{\bf 3D MHD simulation of an eruption in a nested-null topology}. (a-d) logarithm of normalized current density and magnetic field lines (gray represents the pseudostreamer, while yellow and green correspond to the inner fan-spine system). N1 and N2 are inner and outer nulls. BCS and FCS are breakout and flare current sheets (From Wyper et al., in preparation).} 
\label{fig5}
\end{figure*}

\begin{figure*}
\centering{
\includegraphics[width=18cm]{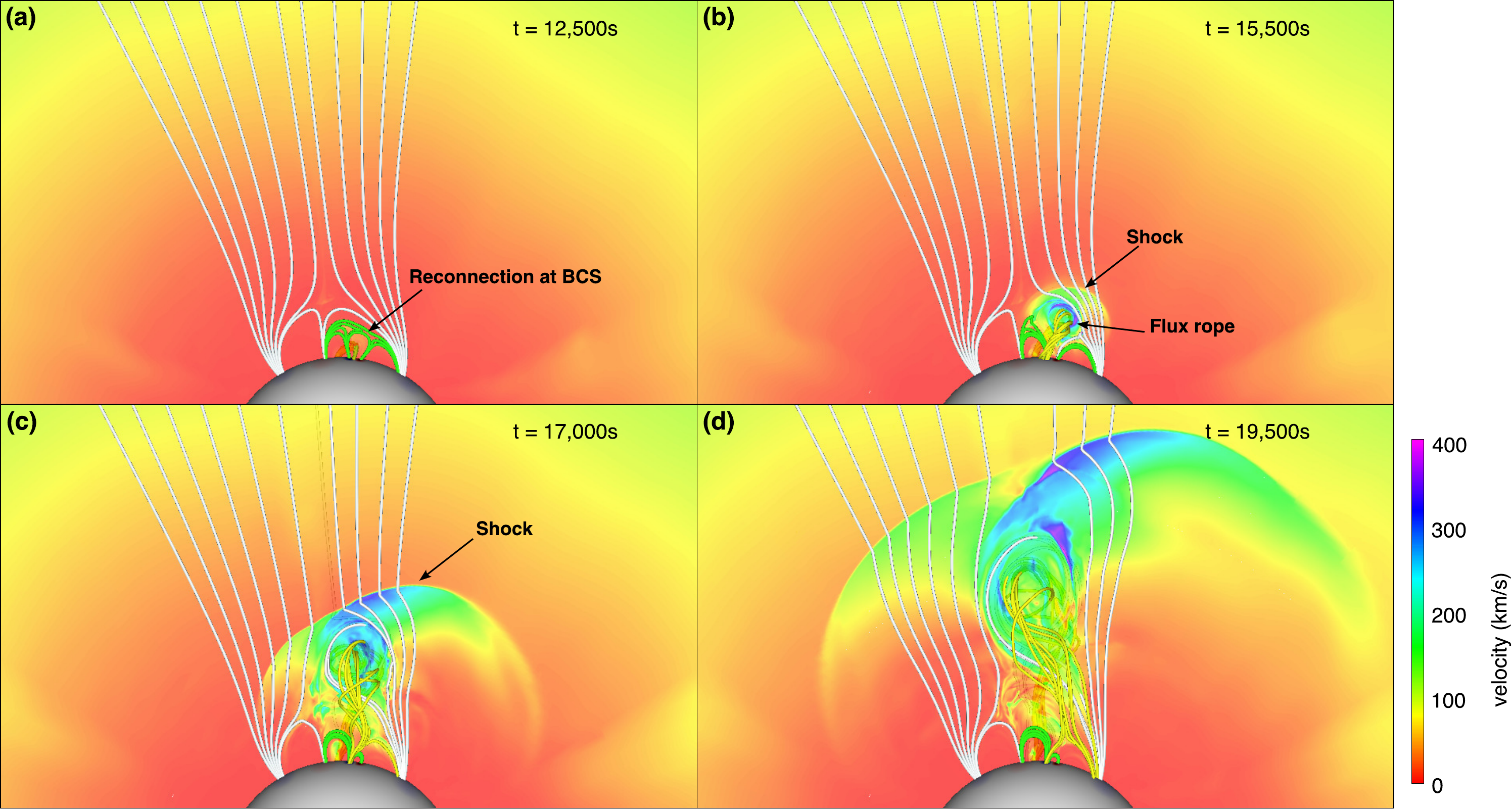}
}
\caption{{\bf Evolution of velocity in the midplane at four times during the simulated eruption}. The color bar represents the velocity scale. The surface gray shading and the field line colors are consistent with those in the previous figure.}  
\label{fig6}
\end{figure*}
\begin{figure*}
\centering{
\includegraphics[width=18cm]{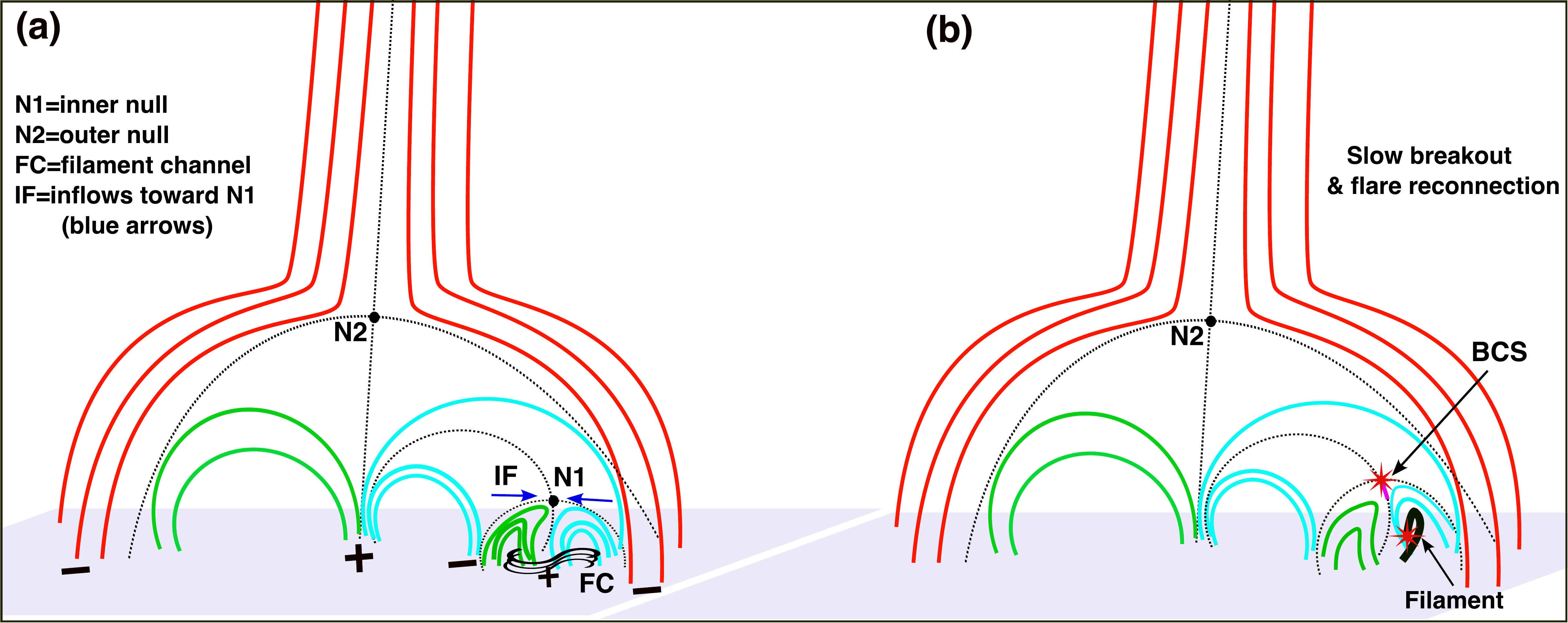}
\includegraphics[width=18cm]{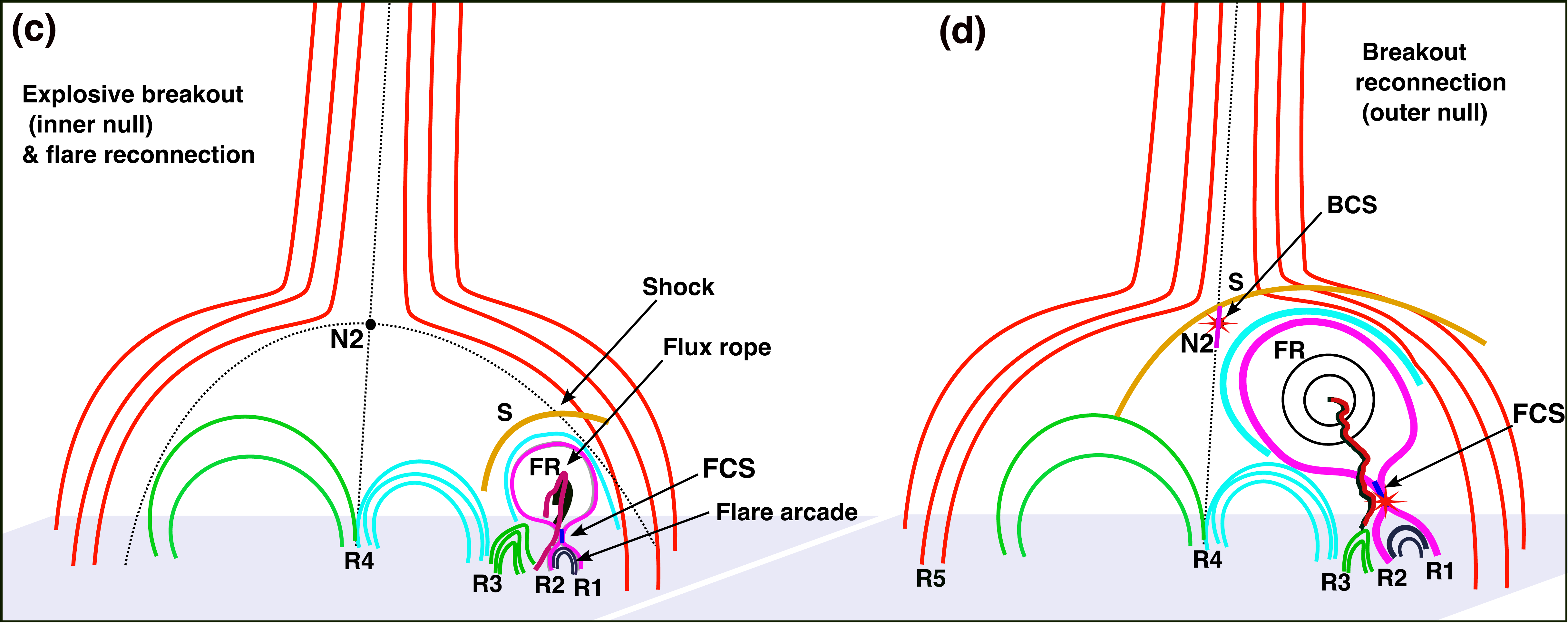}
}
\caption{{\bf Schematic diagram of the energy release process in a nested null-point topology.}
(a) Initial magnetic configuration of the nested fan-spine topology, along with inflows toward inner null N1 during the pre-eruption phase.
(b) Slow magnetic reconnection at the breakout current sheet (BCS) near N1 (formation of arcades below N1). Simultaneous slow flare reconnection (brightenings) below the rising filament-channel core builds the flux rope (FR).
(c) Explosive breakout reconnection near N1 produces ribbons R3 (quasicircular) and R4 (remote). Simultaneous explosive flare reconnection at the flare current sheet (FCS) generates a typical two-ribbon flare.  Quasiperiodic intensity fluctuations at ribbons R3/R4 and  simultaneous metric/decimetric Type III radio bursts. A fast shock wave forms ahead of the erupting flux rope (onset of Type II radio burst).
(d) Interaction of the shock and CME near null point N2 (second breakout reconnection), appearance of a remote ribbon R5. Slow mode waves were observed after the passage of the shock through N2 (MHD wave mode conversion).}

\label{app-fig7}
\end{figure*}

\subsection{Comparison with MHD simulation}
To test our conclusions about the analyzed events, we present an MHD simulation of a successful eruption within a nested-null topology. A detailed analysis and description of the MHD simulation will be presented in a separate paper (Wyper et al., in preparation). We used the three-dimensional Adaptively Refined Magnetohydrodynamics Solver (ARMS; \citet{devore2008}) to model a nested system containing an inner fan-spine system within a pseudostreamer. The properties of the magnetic structures (e.g., null heights and dome widths) and photospheric magnetic flux densities of this model are comparable to those of the observed system. As in our previous work \citep[e.g.][]{wyper2024,karpen2024} the field is constructed through a radial monopole of strength $-0.625$\,G at the bottom boundary combined with subsurface magnetic dipoles which form the nested-null topology. The upper and lower boundaries are open to mass flows (to support an isothermal 1 MK Parker solar wind), while the side boundaries are closed. Additionally, the magnetic field at the lower boundary is line-tied while the side and top boundaries are free-slip. The grid is comprised of $10\times10\times18$ base-level grid blocks (containing $8\times8\times8$ cells), with exponential grid stretching in the radial direction and with four additional levels of adaptive grid refinement employed to give a maximum refinement (at the lower boundary) of $(\Delta r,\Delta \theta,\Delta \phi) \approx (0.026~R_\odot,0.10^\circ,0.10^\circ)$. A similar nested-null topology was previously employed to simulate a failed eruption \citep{karpen2024} but with a stronger background radial field ($-1.25$ G) while keeping everything else the same. Photospheric flows formed and energized the filament channel along a circular PIL in the inner fan-spine system, similar to our previous MHD simulations \citep{wyper2021,karpen2024,wyper2024}. Figure \ref{fig5} and the accompanying animation (Movie S9) illustrate the current density distributions during the eruption. The shearing motions along the PIL accumulate free magnetic energy in the filament channel, driving the slow expansion of the core flux above the PIL in the inner fan-spine system. Slow breakout reconnection at the inner BCS gradually removes the overlying strapping field, while concurrent slow flare reconnection builds the flux rope (shown in yellow) around the filament (Figure \ref{fig5}(a)). The eruption accelerates as explosive breakout reconnection occurs above the flux rope, accompanied simultaneously by fast flare reconnection below it. These processes drive the rapid acceleration of the flux rope and the formation of an MHD shock ahead of it (Figure \ref{fig5}(b)).

When the shock reaches the outer null, N2, it undergoes refraction and leads to the generation of multiple slow-mode waves by mode conversion at N2 \citep[e.g., ][]{McLaughlin2011,kumar2024}. Secondary breakout reconnection occurs at N2 as the flux rope interacts with the open field outside the pseudostreamer (Figure \ref{fig5}(c,d)). Velocity snapshots and the animation (Movie S10) further reveal the flux rope eruption, associated shock formation during breakout reconnection at N1, and the emergence of post-eruption slow-mode waves above N2  (Figure \ref{fig6}).

The observed shock ahead of the flux rope during breakout reconnection at N1 agrees with the MHD simulation. Other key features in the simulation that are qualitatively consistent with the observations and their sequence include the pre-eruption opening at N1, the formation of a hot flux rope, and the development of circular and remote brightenings associated with successive breakout reconnection episodes at N1 and N2. 

Figure \ref{app-fig7} presents a schematic diagram illustrating the energy release process in a nested null-point topology. It highlights the sequence of magnetic reconnection events from slow to explosive breakout reconnection near inner null N1, slow to explosive flare reconnection in the inner system, and the interaction of the erupting flux rope and shock with the outer null N2, leading to formation of multiple ribbons.

\subsection{Implications for confined vs successful eruptions}
All events (E1, E2, E3, and E4) produced successful eruptions. However, prior to events E1 and E3, we also observed confined eruptions of different filaments lying along the circular PIL that did not reach the inner null (N1) and were presumably halted by the overlying arcades. Event E3 and E4 are hybrid jet-CME events, which initially started as a jet. The partial destruction of the filament-carrying flux rope generated a jet, which was followed by its successful eruption (visible in the hot channel), a shock (Type II burst) and a CME, similar to events E1 and E2.

Solar eruptions can either be confined within the lower corona or successfully escape to produce a jet or a CME. The fate of an eruption depends on the complex interplay between the magnetic field configuration, reconnection dynamics, and strength of the overlying strapping field.
In simple fan-spine topologies, the overlying magnetic field strength and connectivity dictate whether an eruption succeeds or fails. If the breakout reconnection is sufficiently strong, it can remove the overlying magnetic field, allowing the flux rope to escape and become a CME \citep{antiochos1999, karpen2012}. However, if the overlying flux remains too strong or reconnection is inefficient due to unfavorable orientation between the flux-rope field and the overlying strapping field, the eruption stalls, leading to a confined event with no CME \citep{kumar2023,kumar2024a}.

In more complex nested-null topologies, eruptions can proceed in multiple stages. The first stage typically involves breakout reconnection at an inner null point, enabling the initial formation of a flare current sheet and flare reconnection, followed by the rapid rise of the growing flux rope with a shock ahead of it. The second stage occurs at the outer null of the pseudostreamer, where additional reconnection determines whether the flux rope can fully escape \citep{karpen2024}. If the outer null reconnection is insufficient due to change of magnetic field orientation of the flux rope (e.g., via kinking), the eruption may remain confined. Conversely, in our cases efficient energy release at both null points as observed here facilitated a successful eruption, leading to a CME with associated shock. 
Here we also observed two confined eruptions in the same nested-null topology where the flux rope and entrained filament rose, kinked, and were unable to reach the inner null. The eruption is confined either due to inadequate breakout reconnection at the inner BCS or because the reconnection above the flux rope halts as the flux rope undergoes rotation (kinking) during the eruption \citep{kumar2022,karpen2024,kumar2025}.
Prior to the successful eruption, observations reveal inflows and expanding core flux converging near the inner null, initiating pre-eruption slow breakout followed by explosive breakout during the eruption of the flux rope.

\subsection{Release and acceleration of energetic particles}
Out of the four homologous events, events E1, E3, E4 (associated with an X-class flare and a halo CME) produced SEPs on March 28-29 (Figure~\ref{app-fig1aa}) and were magnetically connected (Figure \ref{app-fig1} (c)) to the spacecraft (Wind \& SOHO at 1 AU). No SEP was detected at 1 AU from event E2 on March 30 (Figure~\ref{app-fig1aa}). The CME speed in all events was slow to moderate (400–500 \kms) in LASCO coronagraph images; the CMEs in E1, E2, and E4 decelerated in the interplanetary medium (2-15 R$_\odot$), whereas the E3 CME accelerated. All events generated shocks ahead of the erupting flux rope following inner breakout reconnection, along with associated metric Type II bursts in the low corona. All events followed a similar eruption process and exhibited consistent multiwavelength signatures in the EUV and radio observations (see Table 1).

In all events, the observed shock height of about 1.15 R$_\odot$ (0.15 R$_\odot$ above the solar surface) during the flux-rope eruption suggests that particle acceleration may begin in the low corona. The radio observations reveal a metric Type II burst, a signature of shock-accelerated electrons, when the shock passed through the outer null \citep[e.g., ][]{kumar2024a}. This is consistent with our 3D MHD simulation, which reveals the formation of a low coronal shock during inner breakout reconnection and its interaction through the outer null. The rapid acceleration of the erupting flux rope during breakout reconnection via removal of the overlying strapping field likely played a key role in generating the low coronal shock, providing an efficient site for particle acceleration. 

All events produced radio-loud moderate speed CMEs and shocks in the low corona. For event E1, we observed a halo CME (500 \kms) that produced a strong shock capable of accelerating SEPs. 
 The EUV/white-light shock in E1 was faster and wider than those in events E2, E3, and E4.  The shock in event E4 was also faster than the shocks in E2/E3. The gradual SEP event associated with E1 is stronger than that associated with E4.  Event E3 only produced a weak SEP event (only near-relativistic electrons).  In addition, we observed shock–blob interaction in event E1 \citep{kumar2025a} and CME–CME interaction in event E4.
 The shock strength (associated with faster/wider CMEs) appears to be a key factor to determine the intensity of SEPs. Particle acceleration processes involving multiple shocks (e.g., CME–CME interactions where two shocks interact) may be more efficient than shock–blob interactions involving only a single shock. In our case, the CME–CME interaction in E4 did not produce a strong SEP event, likely due to the moderate speeds of the CMEs and their associated shocks.
The absence of energetic protons and electrons at 1 AU during E2 (March 30) may be attributed to unfavorable magnetic connectivity and/or relatively weak CME-driven shocks. However, similar to the other events, radio bursts in the low corona in E2 indicate significant electron acceleration.

Quasiperiodic Type III radio bursts lasting about 10-15 minutes were observed during all events. These bursts originated during breakout reconnection at the inner and outer BCSs, where electrons are most likely accelerated and injected outwards into the corona. The inferred heights of the inner (0.05 R$_\odot$) and outer (0.16 R$_\odot$) nulls provide key constraints on the locations where electrons are efficiently released into interplanetary space. 
NRH radio imaging at 150–327 MHz of event E2 shows quasiperiodic Type III bursts, indicating that the escaping electron beams were produced and/or released by some repetitive process during breakout reconnection at N1. The initial height of the Type III source at 327 MHz is approximately 0.2 R$_\odot$ above the photosphere. The Type II burst images show two sources near the shock flank, with an extension of the sources toward the shock front. This event provides a clear example of two Types of radio signatures associated with breakout reconnection \citep[e.g., ][]{aurass2013,kumar2013a}.
The recurrent injection of electrons during magnetic reconnection at N1/N2 may serve as a seed population for shock acceleration during the initial stage of the SEP event detected at 1 AU.  The dynamic radio spectra in the metric/decimetric range (low corona) show cotemporal Type II and Type IIIs, suggesting that both processes (injection and shock acceleration) occurred simultaneously. This implies that the initial stage of the SEP event detected at 1 AU was likely influenced by both mechanisms. Interplanetary decametric–hectometric (DH) Type III bursts were observed in all events as an extension of low coronal decimetric/metric Type III bursts. None of the events exhibited interplanetary Type II bursts, except during the shock interaction with a slowly moving blob in Event E1. However, faint shock fronts were observed in all events in the coronagraph images. 

For the event E1, we estimated the electron release time at the sun using the velocity dispersion analysis method. The electron release time was compared with the timing of decimetric/metric Type IIIs and metric Type II radio emissions. This release time (17:48 UT$\pm$2.6 min) was close to the onset of explosive breakout reconnection at the inner/outer null (17:45-17:53 UT). The shock appeared ahead of the flux rope at 17:47 UT (projected height$\approx$0.15 R$_\odot$ above the solar surface) during the inner null reconnection.  A close temporal association between electron release and Type III emission suggests that electrons were primarily injected recurrently during the breakout reconnection at the inner/outer null. The proton release time was delayed, on the other hand, consistent with gradual acceleration at the shock front. It is likely that the recurrent injection of particles from the inner/outer nulls may contribute to the seed population during the initial stage (10-15 minutes) of the SEP event detected at 1 AU. Later on, shock acceleration is the most efficient mechanism to produce the gradual SEP event. Our findings highlight the complex interplay between breakout reconnection, shock formation, and associated particle acceleration.

We observed a slowly moving blob in the LASCO C2 coronagraph images prior to the CME and detected radio signatures of its interaction with the shock at about 7 R$_\odot$. This interaction may enhance particle acceleration efficiency as the shock propagates through the blob. Subsequently, the blob merged into the CME. The observed radio enhancement (a Type II burst in the 2–3 MHz band of the radio dynamic spectrum) is similar to those seen in larger CME-CME interactions \citep{gopalswamy2001,gopalswamy2002}. The estimated source speed, derived from the drift rate of the Type II burst, is consistent with the shock speed derived from the white-light images. A detailed analysis of the radio signatures associated with the shock-blob interaction and merger is presented in a separate paper \citep{kumar2025a}.

\section{CONCLUSIONS}
For the sequence of events analyzed here, we demonstrated that breakout reconnection near the inner null point is responsible for the onset of the CMEs, the injection of energetic electrons that produce remote ribbons via recurrent precipitation of downward-moving electron beams, and simultaneous electron beams escaping outward into the interplanetary medium emitting quasiperiodic Type III bursts. Breakout reconnection provides positive feedback to flux ropes during eruptions by continuously reducing the overlying magnetic constraints and facilitating further expansion and acceleration of the erupting flux rope. The ongoing flare reconnection in the filament channel builds the flux rope observed in hot channels during the eruption. 
Flare reconnection below the flux rope produces a typical two-ribbon flare (R1, R2), along with a circular ribbon (R3) and remote ribbons (R4, R5) through breakout (interchange) reconnection at the inner and outer null points. Therefore, successful eruptions in nested fan-spine topologies result in five-ribbon flares. The EUV/radio observations provide evidence of quasiperiodic injection of electron beams (metric/decimetric radio bursts) during the breakout reconnection.
Strong shocks appeared in the low corona ahead of the flux rope during the explosive breakout reconnection near the inner null. The rapid acceleration of the flux rope during breakout reconnection led to the formation of a shock in the low corona. These shocks are primarily responsible for the acceleration of energetic particles in the low corona and interplanetary medium and the associated  Type II radio bursts, leading to the production of gradual SEPs detected in the heliosphere. 
This study highlights the role of nested-null topologies in producing some solar eruptions, and demonstrates that two successive breakout reconnections at inner and outer nulls are key to the initiation of CMEs, shocks, and associated 
acceleration of SEPs. By examining both pre-eruption dynamics and the explosive reconnection events, it provides new insights into how energetic particles are released and accelerated, advancing our understanding on the solar origin of SEPs. These results have broad implications for understanding radio emissions and particle acceleration in null-point topologies, which are ubiquitous on the Sun \citep{kumar2019a,kumar2021,kumar2022} and play key roles in producing the wide range of eruptions (i.e., jets and CMEs). These CMEs, shocks, and associated SEPs are particularly important for space weather.
 Future studies incorporating detailed MHD simulations, as well as multiwavelength and multipoint observations from Solar Orbiter and Parker Solar Probe \citep[e.g., ][]{lario2024}, will be essential for refining our understanding of these processes.\\
 \\

\noindent
{\bf {Acknowledgements}}\\
We are grateful to the reviewer for constructive comments/suggestions, which have improved the paper.
SDO is a mission for NASA's Living With a Star (LWS) program. The LASCO CME catalog is generated and maintained at the CDAW Data Center by NASA and The Catholic University of America in cooperation with the Naval Research Laboratory. SOHO is a project of international cooperation between ESA and NASA. We thank R. Lin and S. Bale (UC Berkeley) and CDAWeb for providing the SEP data. This research was supported by NSF SHINE grant (\#2229336), NASA’s Heliophysics Supporting Research (\#80NSSC24K0264), and Guest Investigator (\#80NSSC25K7679) programs. 
The computations were sponsored by allocations on Discover at NASA's Center for Climate Simulation and on the DiRAC Data Analytic system at the University of Cambridge, operated by the University of Cambridge High Performance Computing Service on behalf of the STFC DiRAC HPC Facility (www.dirac.ac.uk) and funded by BIS National E-infrastructure capital grant (ST/K001590/1), STFC capital grants ST/H008861/1 and ST/H00887X/1, and STFC DiRAC Operations grant ST/K00333X/1. DiRAC is part of the National E-Infrastructure. PW was supported by an STFC (UK) consortium grant ST/W00108X/1 and a Leverhulme Trust Research Project grant.
We thank the RSDB service at LESIA/USN (Observatoire de Paris) for making the NRH/ORFEES/NDA data available. We equally thank the radio monitoring service at LESIA (Observatoire de Paris) for providing value-added data used for this study. Wavelet software was provided by C. Torrence and G. Compo, and is available at http://paos.colorado.edu/research/wavelets/.\\
\\

\bibliographystyle{aasjournal}
\bibliography{reference.bib}

\clearpage
\appendix
\counterwithin{figure}{section}
\section{Supplementary figures}
\begin{figure*}[htp]
\centering{
\includegraphics[width=14cm]{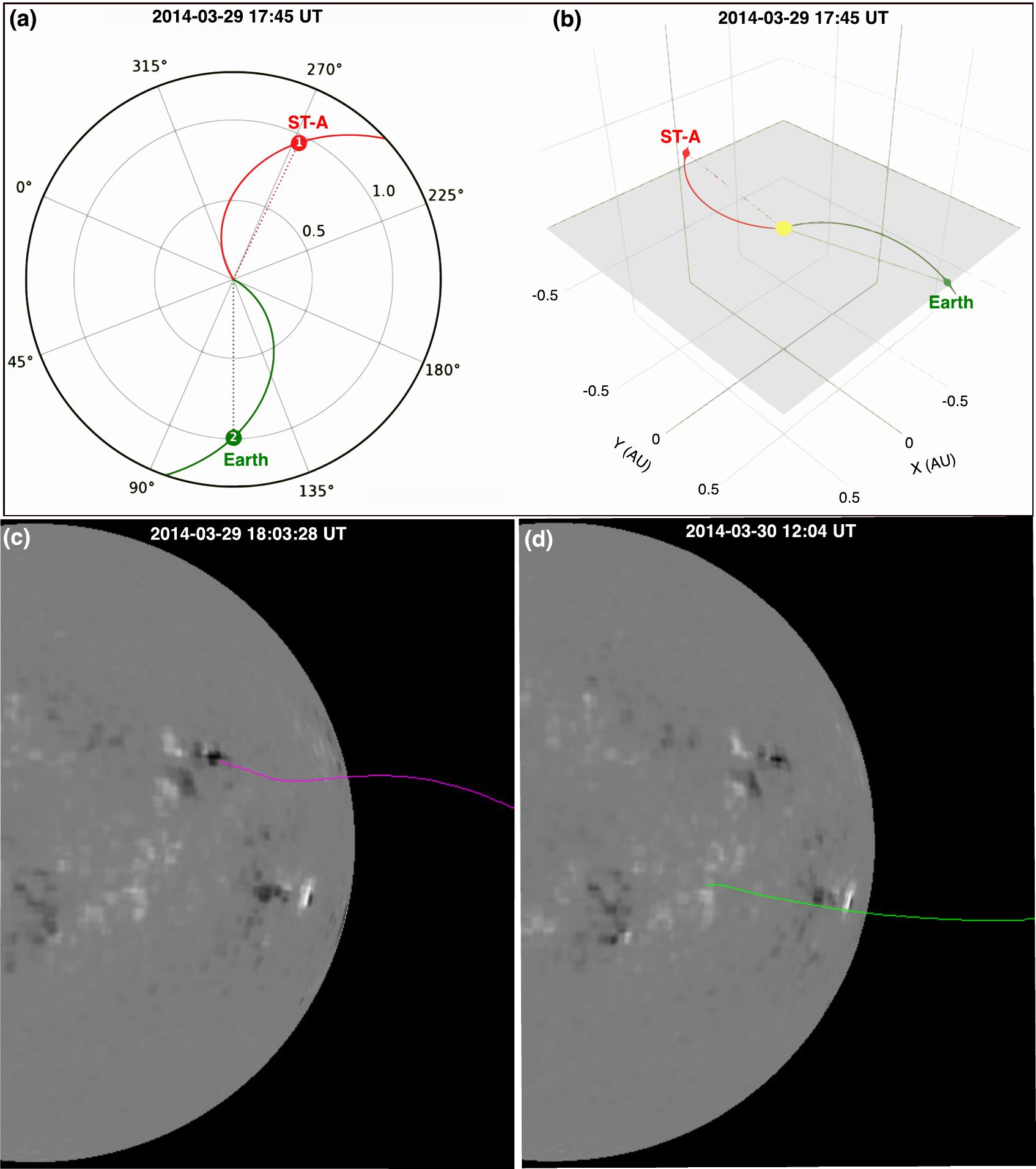}
}
\caption{{\bf Magnetic connectivity.} (a,b) Magnetic connectivity maps (two different views) for Earth and STEREO-A on March 29, 2014, at 17:45 UT (Event E1). (c,d) The footpoint connectivity of the Earth-directed field lines (from 2.5 R$_{\odot}$ back to the solar surface) using the PFSS extrapolation for March 29 (pink) and March 30 (green).} 
\label{app-fig1}
\end{figure*}

\begin{figure*}[htp]
\centering{
\includegraphics[width=13cm]{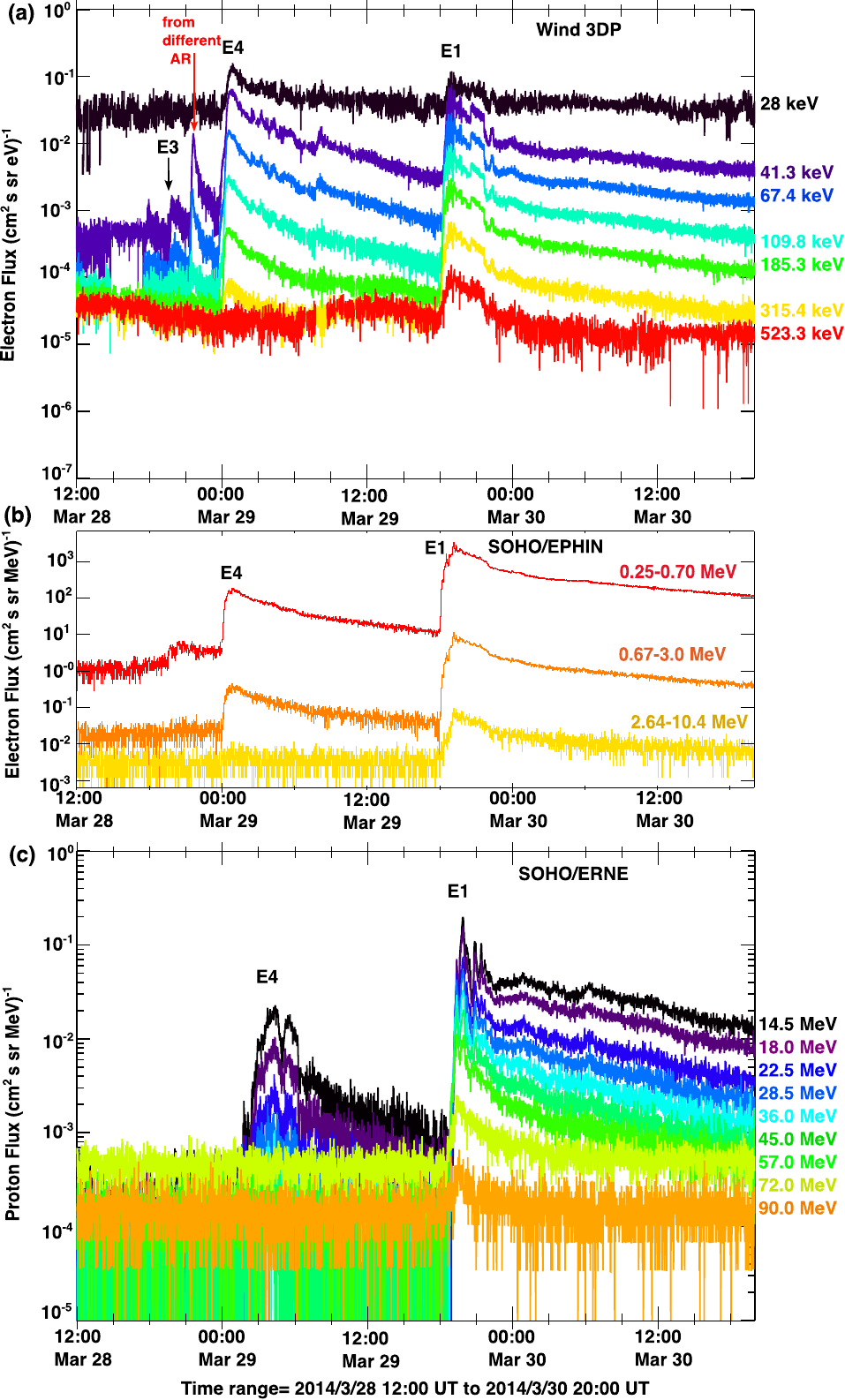}
}
\caption{{\bf Solar energetic particles (electrons/protons) during all events (E1-E4).} (a) Electron flux (12-s cadence) measured by Wind 3DP at different energies ranging from 28-523 keV. (b) Electron flux (1 min cadence) detected by SOHO/EPHIN at different energies ranging from 0.25-10 MeV. (c) Proton flux (1 min cadence) detected by SOHO/ERNE at different energies ranging from 14.5-90 MeV. No SEPs were detected from Event E2 on March 30.}  
\label{app-fig1aa}
\end{figure*}

This appendix contains additional figures to support the results described above. 
Figure \ref{app-fig1} shows Earth and STEREO-A's magnetic connectivity using the Solar-MACH (Solar Magnetic Connection Haus: \citealt{gieseler2023}) tool. The magnetic footpoint's Carrington longitude was 175.4$^\circ$, and the Carrington latitude was -6.7$^\circ$ for the Parker spiral connecting Earth (V$_{sw}$=400 \kms) on March 29 (17:45, 18:00 UT). The magnetic footpoint connecting to Earth through the Parker spiral on March 30 (12:00 UT) had a Carrington longitude of 165.4$^\circ$ and a Carrington latitude of -6.6$^\circ$. We used PFSS extrapolation to determine the footpoint connectivity from 2.5 R$_{\odot}$ down to the solar surface (Figure \ref{app-fig1}(c,d)). Note that the AR NOAA 12017 was magnetically connected to Earth (pink field line) on March 29, whereas it was not connected on March 30 (green field line). The SEPs (electrons and protons) associated with the homologous flares/eruptions discussed in this paper are shown in Figure \ref{app-fig1aa}. Events E1 and E4 both produced gradual SEP events (electrons and protons), whereas Event E3 produced a weak SEP event involving only near-relativistic electrons (41.3-67.4 keV). 

\begin{figure*}[htp]
\centering{
\includegraphics[width=18cm]{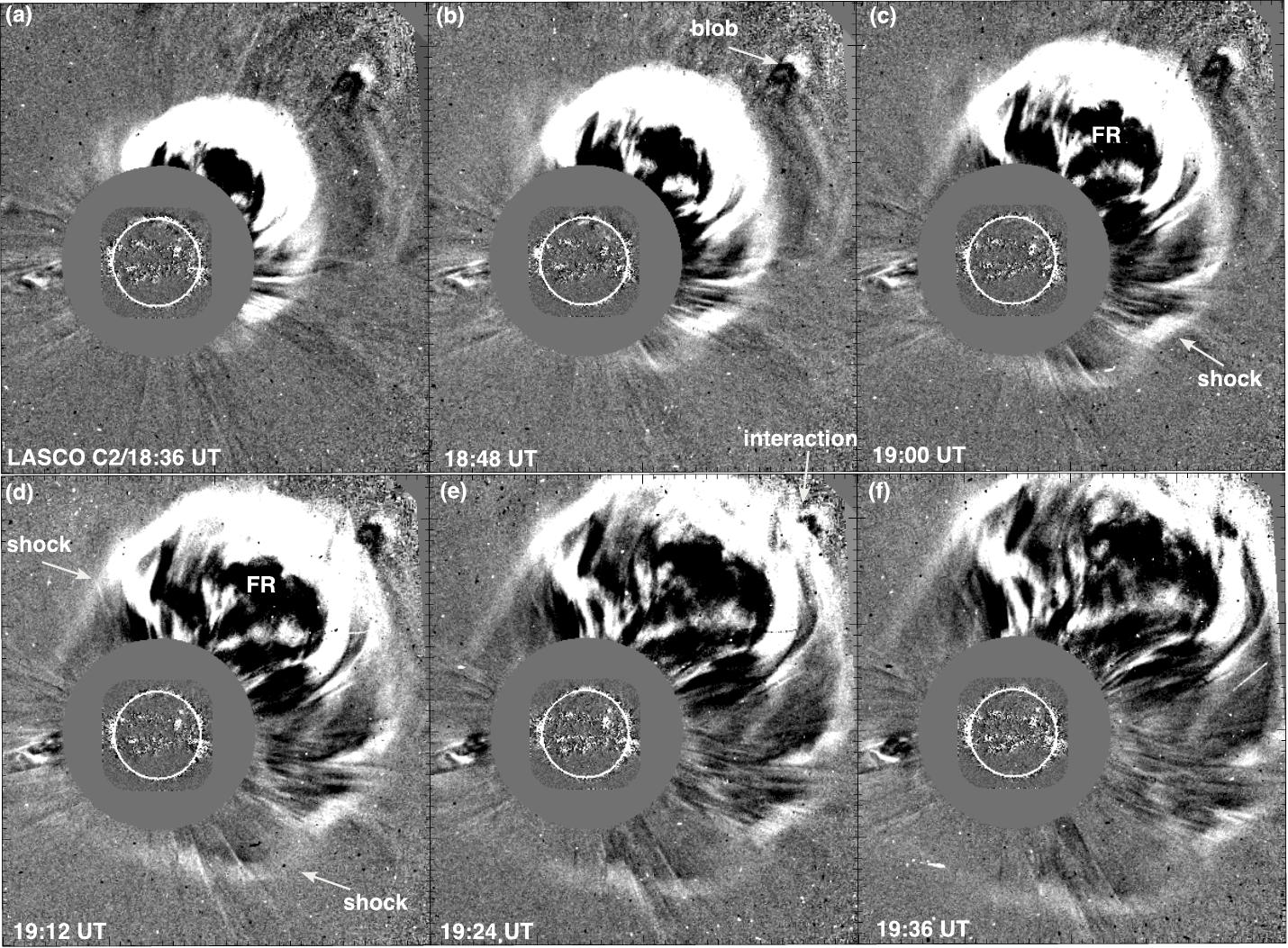}
}
\caption{{\bf LASCO C2 observations of the CME and slow blob for Event E1 on March 29, 2014.} (a-f) LASCO C2 (2-6 R$_\odot$) running-difference images of the CME and the associated shock (marked by arrows). FR represents the flux rope. Panel (e) illustrates the interaction between the CME shock and the blob.} 
\label{app-fig1a}
\end{figure*}

\begin{figure*}[htp]
\centering{
\includegraphics[width=18cm]{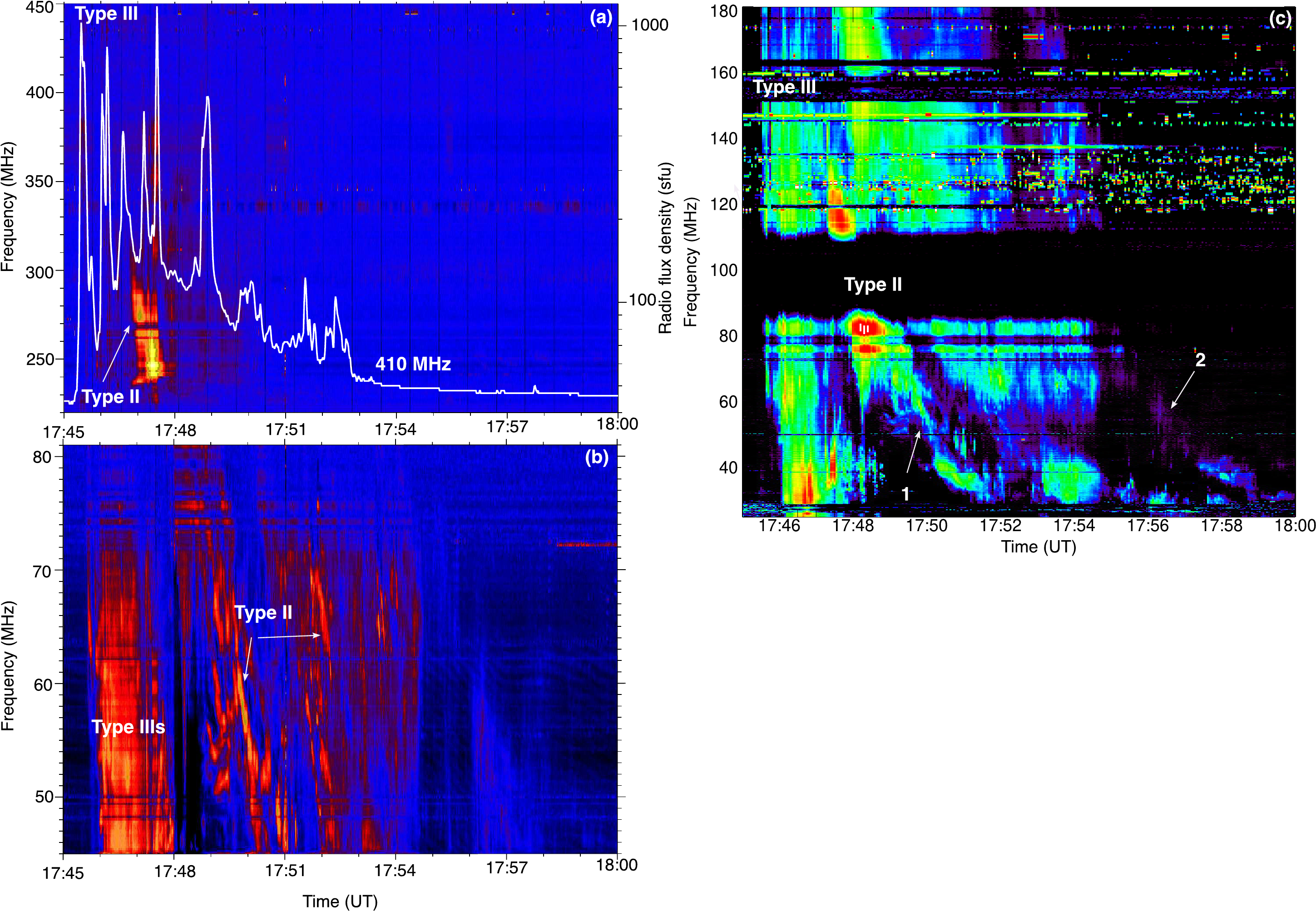}
}
\caption{{\bf Dynamic radio spectra for Event E1}. (a,b) Dynamic radio spectra from e-Callisto (Roswell Observatory, New Mexico). The white-line plots the 410 MHz radio flux density profile (1-second cadence) from the RSTN Sagamore Hill Radio Observatory (scale is on the right Y axis). (c) Dynamic radio spectrum from the RSTN Sagamore Hill Radio Observatory in the 25–180 MHz range. The Type II bursts are indicated by 1 and 2.} 
\label{app-fig1b}
\end{figure*}

\begin{figure*}[htp]
\centering{
\includegraphics[width=18cm]{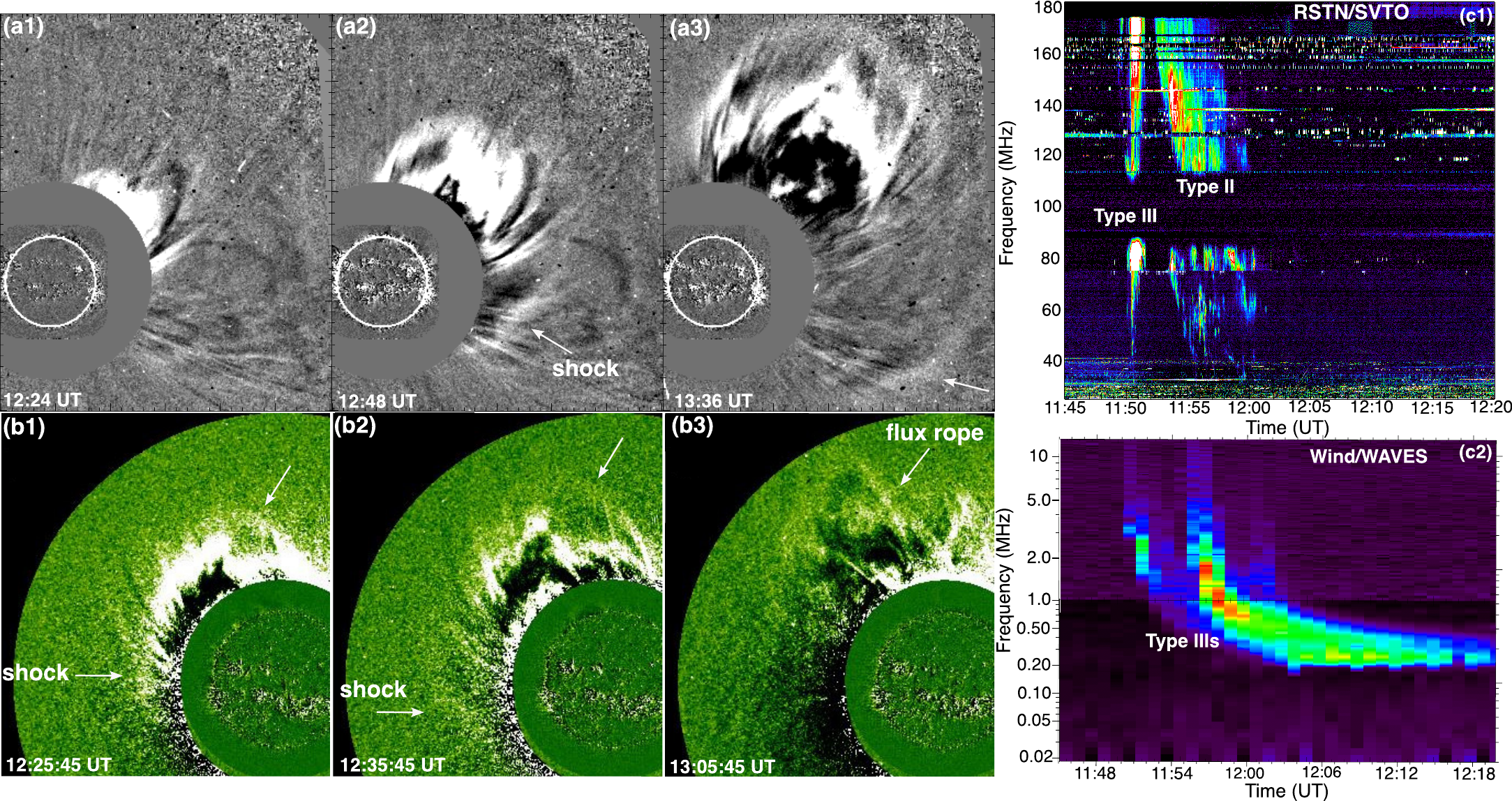}
}
\caption{{\bf LASCO C2 and STEREO-B COR1 observations of the CME for Event E2 on March 30, 2014.} (a1-a3,b1-b3) LASCO C2 (2-6 R$_\odot$)  and STEREO-B COR1 (1.4-4 R$_\odot$) running-difference images of the CME and the associated shock (marked by arrows). The flux rope structure is evident in panel (b3). (c1,c2) Dynamic radio spectra from the RSTN San Vito station in the 25–180 MHz range and from Wind/WAVES (0.02–13.82 MHz).} 
\label{app-fig2}
\end{figure*}

\begin{figure*}[htp]
\centering{
\includegraphics[width=16cm]{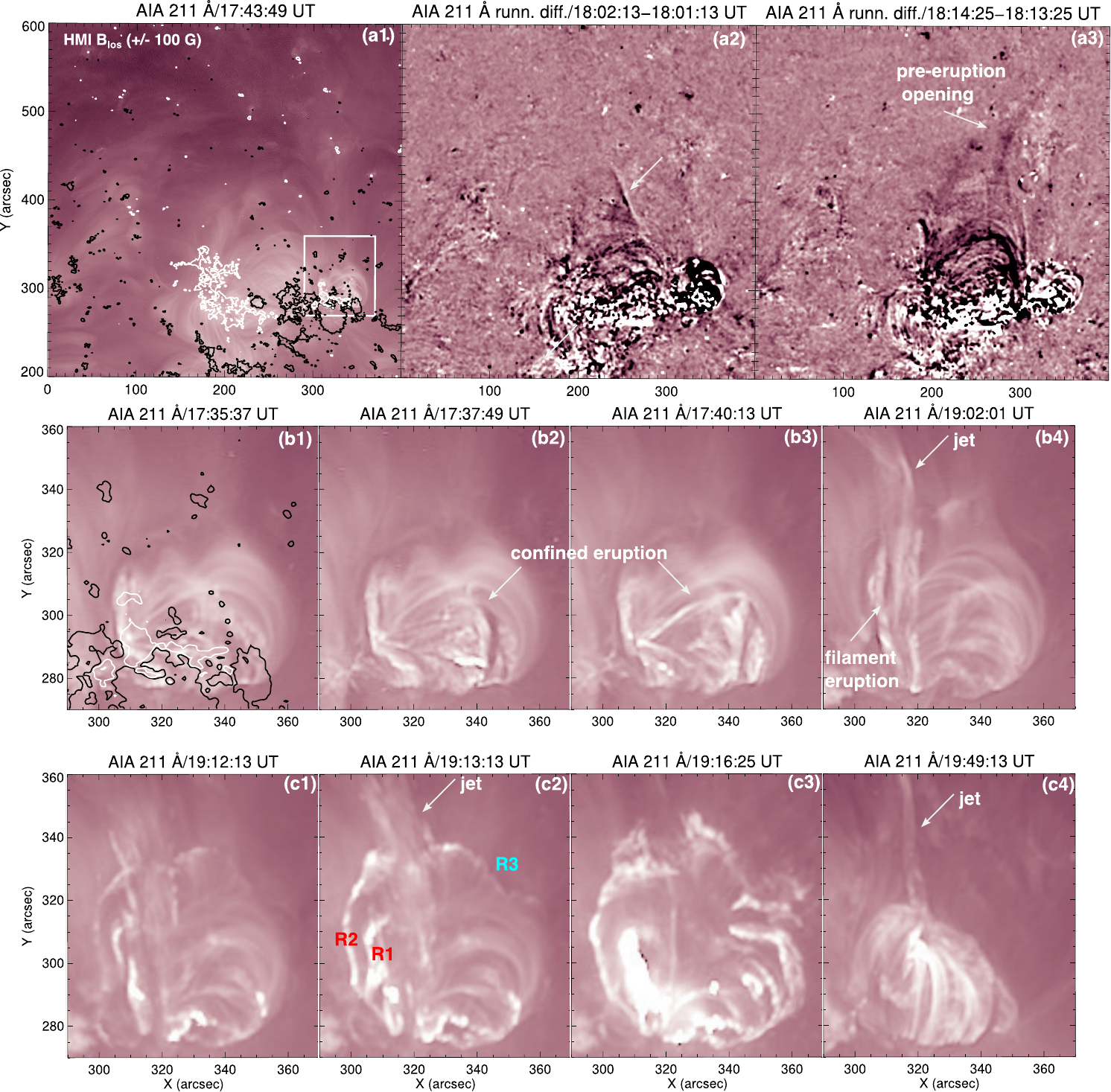}
}
\caption{{\bf SDO/AIA observations of Event E3 on March 28, 2014.} (a1-a3) AIA 211~{\AA} intensity and running-difference images during pre-eruption opening (marked by arrows) above the right lobe of the pseudostreamer. The first image is overlaid with cotemporal HMI magnetogram contours of positive (white) and negative polarities, scaled at $\pm$100 G. (b1-b4,c1-c4) AIA 211~{\AA} images of the region outlined by the white box in (a1) illustrating a confined filament eruption followed by a successful eruption associated with a recurring jet and an M2.0 flare.} 
\label{app-fig3}
\end{figure*}

\begin{figure*}[htp]
\centering{
\includegraphics[width=18cm]{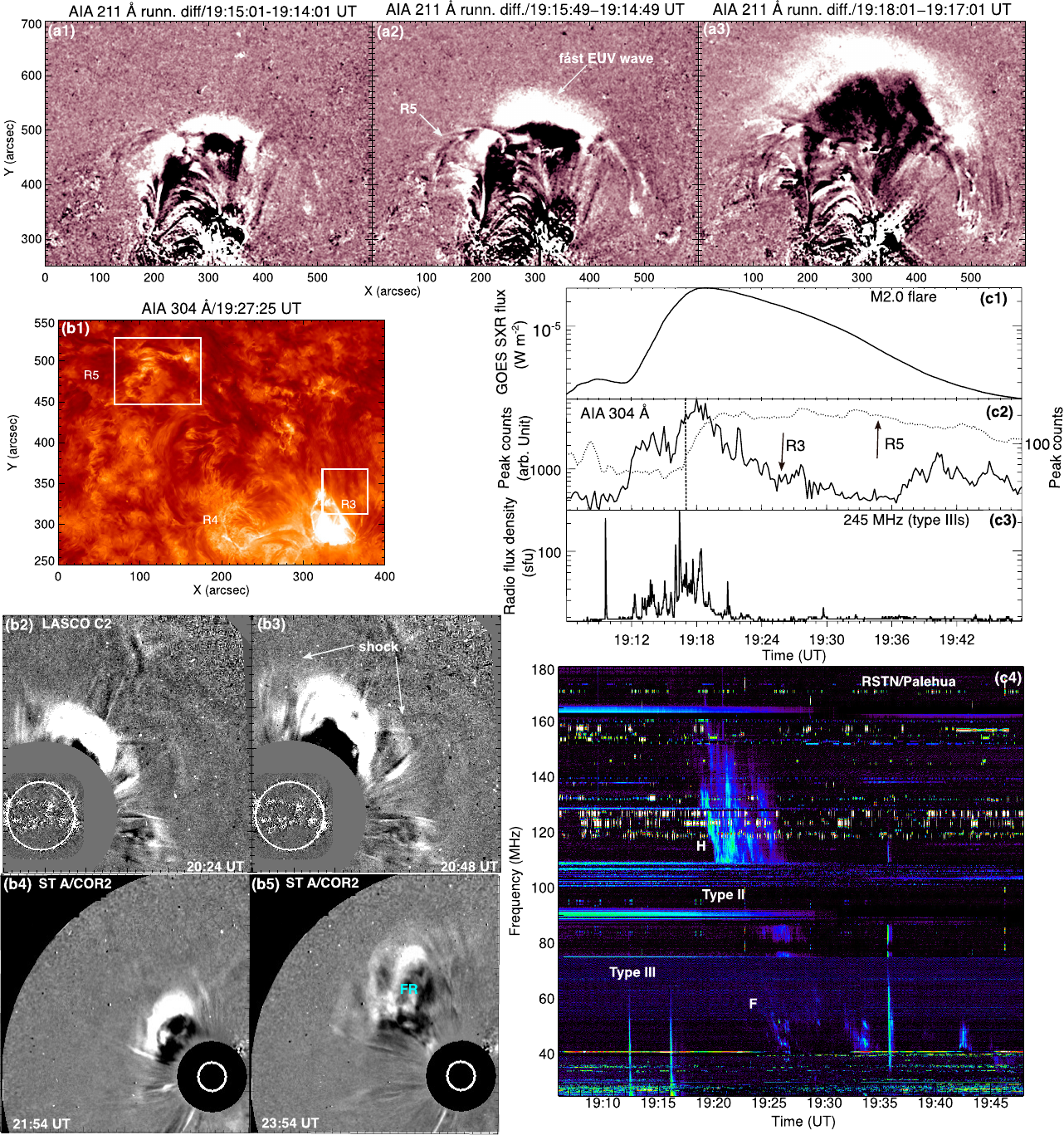}
}
\caption{{\bf Fast EUV wave, radio bursts, and associated CME for Event E3 on March 28, 2014.} (a1-a3) AIA 211~{\AA} running-difference images showing a large-scale EUV wave associated with the filament eruption and jet. (b1) AIA 304~{\AA} image displaying remote ribbons R4 and R5 during the eruption. (b1-b4) LASCO C2 (2-6 R$_{\odot}$) and STEREO-A COR2 (2-15 R$_{\odot}$) images depicting the CME and shock. FR is the flux rope. (c1,c2,c3) GOES soft X-ray flux profile in the 1–8~{\AA} channel, AIA 304~{\AA} peak counts extracted from region R3 and R5 (within the box marked in (b1)), and the radio flux density profile (1-s cadence) at 245 MHz from the RSTN Sagamore Hill Radio Observatory. The vertical dashed line indicates the onset of energy release at N2. (c4) Dynamic radio spectrum in 25-180 MHz from the RSTN Palehua Radio Observatory.} 
\label{app-fig4}
\end{figure*}

Figures \ref{app-fig1a}, \ref{app-fig1b}, and \ref{app-fig2} display the CME and associated radio signatures for event E1 and E2. Figures \ref{app-fig3} and \ref{app-fig4} illustrate the multiwavelength analysis (EUV and white-light images along with radio signatures) for event E3. 

\begin{figure*}[htp]
\centering{
\includegraphics[width=18cm]{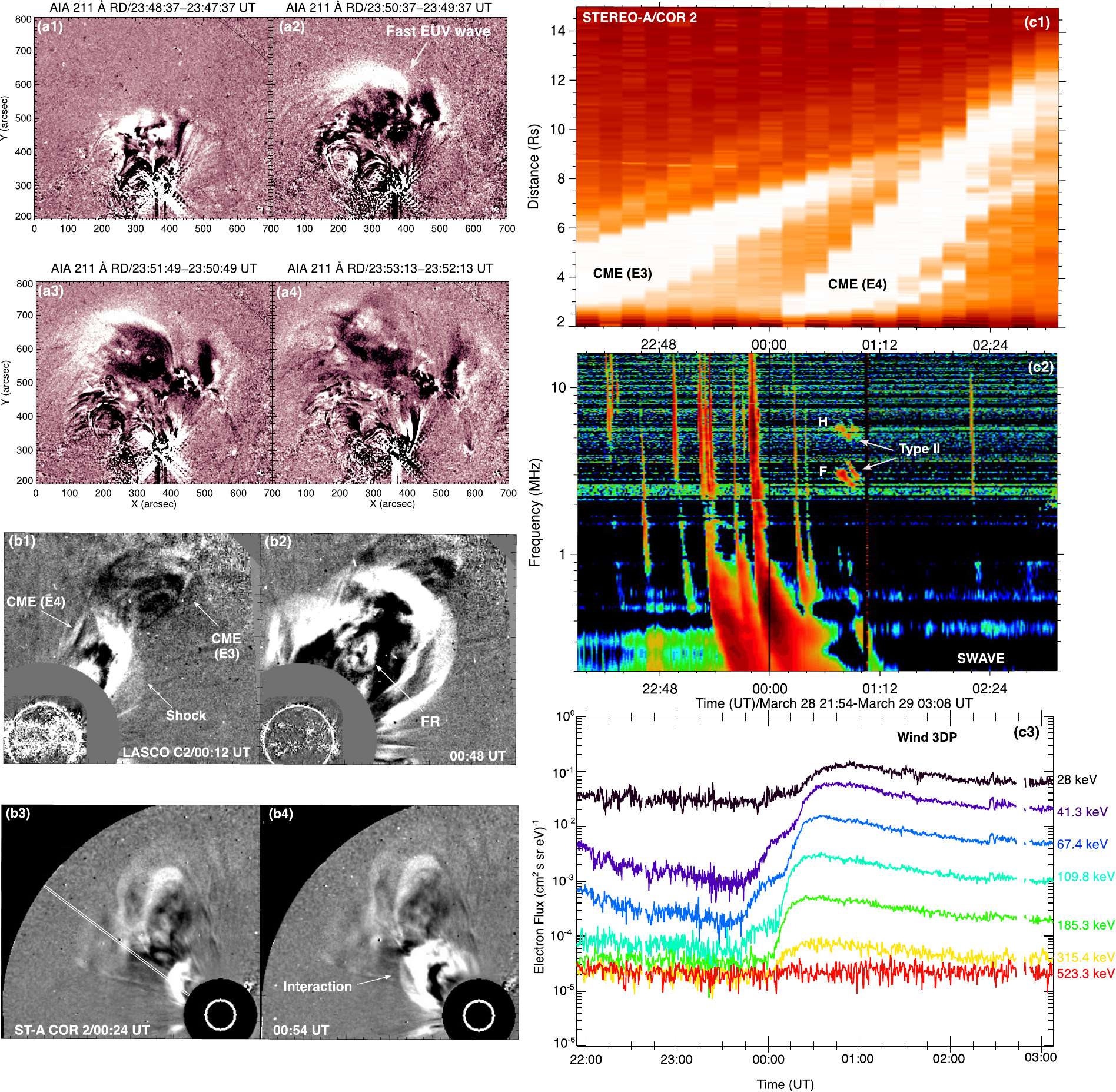}
}
\caption{{\bf Fast EUV wave, CME, and associated SEPs for Event E4 on March 28, 2014.} (a1-a4) AIA 211~{\AA} running-difference images showing a large-scale EUV wave associated with a filament eruption. (b1-b4) LASCO C2 (2-6 R$_{\odot}$) and STEREO-A COR2 (2-15 R$_{\odot}$) images depicting the CME and shock. FR is the flux rope. CME (E3) is the preceding CME from the earlier eruption E3. (c1) Time–distance intensity plot along the slit (marked in panel b3) using STEREO-A/COR2 white-light images. (c2) Dynamic radio spectrum in 0.2-16 MHz from STEREO-A/WAVES. F and H indicate the fundamental and harmonic bands of the Type II burst. (c3) Electron flux (12-s cadence) measured by Wind 3DP at different energies ranging from 28-523 keV.}
\label{app-fignn}
\end{figure*}

Events E3 and E4 originated from eruptions from the same PIL, separated by $\approx$4.5 hours. The CME/shock associated with E4 was faster than that of E3 and interacted with E3 at around 7~$R_\odot$ (Figure \ref{app-fignn} (b1-b4, c1). An interplanetary Type II radio burst was observed at 00:40-00:56 UT when the shock passes through the plasma sheet and flux-rope of the preceding CME (Figure \ref{app-fignn} (c2)). A clear band-splitting feature is observed in both the fundamental and harmonic emissions of the Type II burst. A Type III radio burst (and associated enhancement in electron flux at 1 AU) was detected at about at 02:12 UT, during the merging of CMEs (Figure \ref{app-fignn} (c2,c3)). Both CMEs produced a single merged structure.  CME–CME interactions can enhance particle acceleration as the shock from the trailing CME propagates through the dense plasma (seed population) of the preceding CME, where the Alfvén speed is relatively low \citep{gopalswamy2001,gopalswamy2002}. This causes the shock to strengthen while traversing the dense plasma sheet or flux rope, thereby providing favorable conditions for efficient particle acceleration. Since both CMEs originate from the same PIL of the active region, their similar twist is likely favorable for magnetic reconnection between the two flux ropes (merging) from E3 and E4.

\begin{figure*}[htp]
\centering{
\includegraphics[width=18cm]{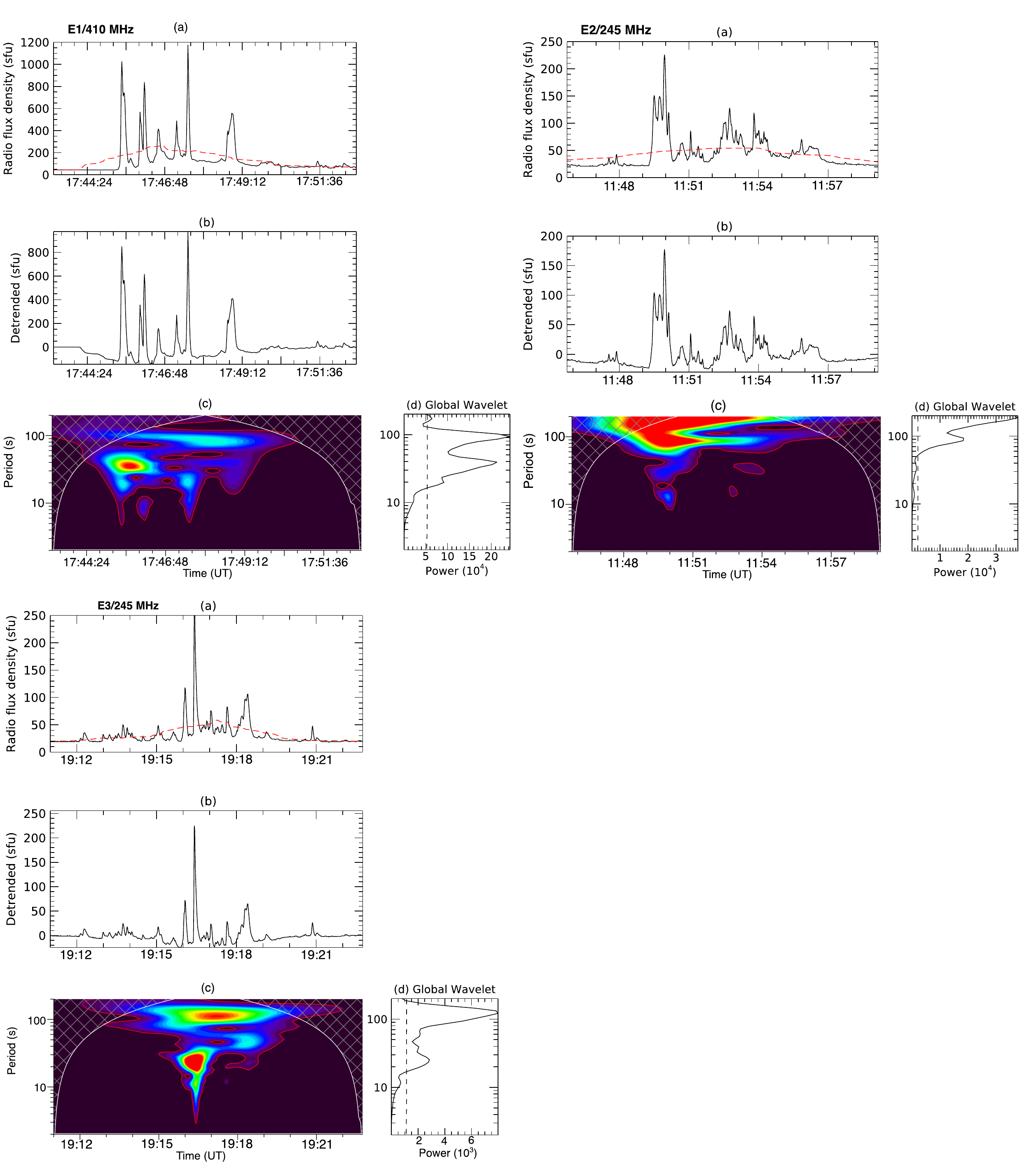}
}
\caption{{\bf Quasiperiodic Type IIIs during breakout reconnection near inner/outer nulls.} (a-d) The wavelet analysis of the radio flux density from RSTN Sagamore Hill observatory for event E1. {\it left:} (a) Radio flux density (410 MHz) in solar flux units (1 sfu=10$^{-22}$ W m$^{-2}$ Hz$^{-1}$) vs time (UT). (b) The detrended light curve after subtracting the red trend shown in (a) from the original light curve. (c) Wavelet power spectrum of the detrended signal. Red contours outline the 95\% significance level. (d) Global wavelet power spectrum. The dashed line is the 95\% global confidence level. {\it right and bottom left:} (a-d) The same plots for events E2 and E3.
} 
\label{app-fig5}
\end{figure*}

\begin{figure*}[htp]
\centering{
\includegraphics[width=14cm]{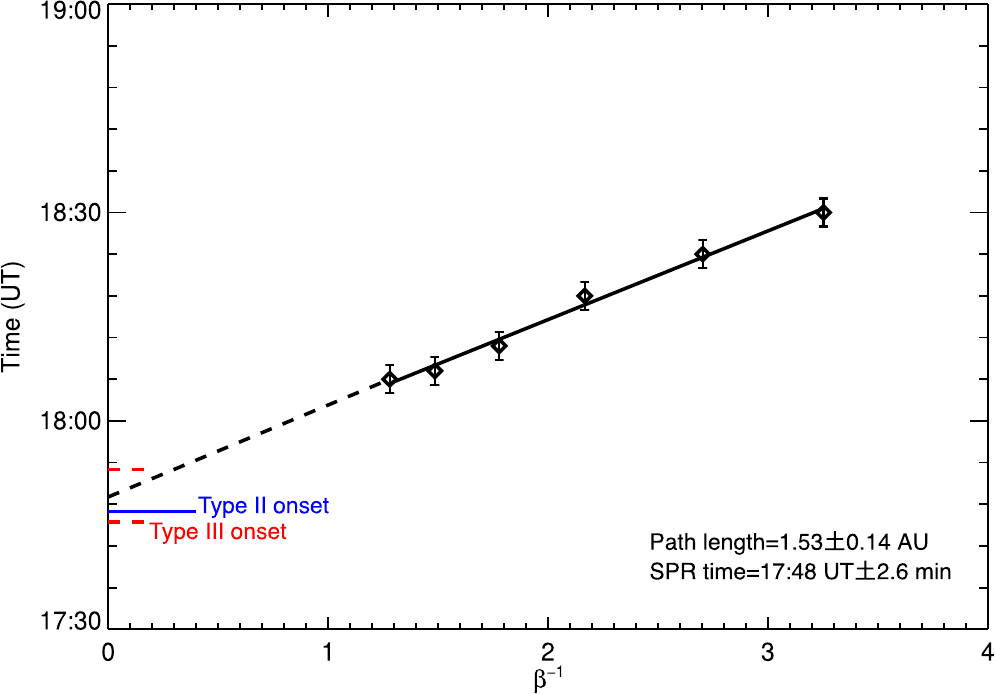}
}
\caption{{\bf Velocity dispersion analysis for the onset of SEPs for E1 on March 29, 2014}. Inverse beta vs onset time plot for electrons (26–307 keV) detected by Wind 3DP at 1 AU. The red dashed lines indicate the onset and end times of decimetric/metric Type III bursts, while the blue line marks the onset time of the metric Type II radio burst.
} 
\label{app-fig6}
\end{figure*}

Figure \ref{app-fig5} depicts the results of our wavelet analysis for the Type III bursts detected by RSTN radio stations. Figure \ref{app-fig6} shows the result of VDA analysis for the estimation of SPR time.



\clearpage
{\small
\begin{longtable*}{c c c c c c }
\caption{Summary of homologous flares and CMEs in NOAA AR 12017} \\
\hline \\
\label{tab1}
Event/Date   &Flare class     & CME properties                    &Type IIIs     & Type II         &SEP (1 AU)  \\
(2014)  &(start, peak, end in UT)   &LASCO C2                  &start/end (in UT)     &start/end (in UT)        &start/end (in UT)   \\
  &   &speed (\kms)         &            &speed (\kms)      &  (electrons/protons) \\ 
  &   &width (deg)          &            &      &   \\ 
\hline
 E1  & X1.0             &18:12:05         &17:45-17:53  &17:47-18:01    & 18:00-23:00    \\  
March 29  &17:35, 17:48,17:54  &528 (halo)    &            &1194,1488   &     \\
&              &360$^{\circ}$     &         &   &     \\ 
    &              &     &         &   &     \\ 
 E2  &M2.1                &12:24:05      &11:49-11:51    &11:52-12:01   & no SEP  \\  
  March 30  &11:48, 11:55, 12:02   &487 (partial halo)     &        &879  & \\
  &              &192$^{\circ}$     &         &   &     \\ 

   &              &     &         &   &     \\ 
 E3  &M2.0                &20:00      &19:09-19:24    &19:18-19:43   & 19:40-21:00 \\  
  March 28  &19:04, 19:18, 19:27   &420      &            &528         &(only electrons)\\
  &              &103$^{\circ}$     &         &   &     \\ 

   &              &     &         &   &     \\ 
 E4  &M2.6                &23:48:07      &23:46-23:52    &23:50-00:03   & 23:56-06:00  \\  
  March 28  &23:44, 23:51, 23:58   &514      &            &857               &  \\
  &              &138$^{\circ}$     &         &   &     \\ 
  
 \\      
 \hline
\end{longtable*}
\small
\noindent
}
\section{Supplementary Materials}
This section contains supplementary movies to support the results. All supplementary movies are available in the Zenodo repository at doi:\href{https://doi.org/10.5281/zenodo.17116252}{10.5281/zenodo.17116252}. \\
{\bf Movie S1}: An animation of the AIA 171~{\AA} and 94~{\AA} images (Event E1), along with the TD plot along slices R1S1 and R2S2, during the pre-eruption phase (Figure \ref{fig1a}). The animation runs from 16:01:37 UT to 17:57:11 UT. Its real-time duration is 16.5 s. \\
{\bf Movie S2}: An animation of the AIA 171, 211, 131, and 94~{\AA} images (Event E1), along with the AIA 304~{\AA} peak counts (extracted from the boxed area in ribbon R4) and Type III radio bursts (410 MHz), during the eruption (Figure \ref{fig1b}). The animation runs from 17:01:49 UT to 18:15:25 UT. Its real-time duration is 9.2 s. \\
{\bf Movie S3}: An animation of the AIA 211~{\AA} intensity and running-difference images (Figure \ref{fig1c}) during the eruption and the formation of fast and slow EUV waves (Event E1). The animation runs from 17:02:49 UT to 18:17:25 UT. Its real-time duration is 7.4 s.\\
{\bf Movie S4}: An animation of the AIA 171, 211, 131, and 94~{\AA} images (Event E2), along with the AIA 211~{\AA} peak counts (extracted from the boxed area in ribbon R3/R4) and Type III radio bursts (245 MHz), during the eruption (Figure \ref{fig-aia}). The animation runs from 11:03:01 UT to 12:07:37 UT. Its real-time duration is 10.8 s. \\
{\bf Movie S5}: An animation of the AIA 211~{\AA} intensity and running-difference images (Figure \ref{ribb-aia}) during the eruption and the formation of fast and slow EUV waves (Event E2). The animation runs from 11:02:49 UT to 12:07:37 UT. Its real-time duration is 8.1 s.\\
{\bf Movie S6}: An animation of the AIA 211~{\AA} intensity and running-difference images (Figures \ref{fig3} and \ref{fig4}) along with NRH radio images (150 and 327 MHz) and dynamic spectra (Event E2). The animation runs from 11:31:13 UT to 12:03:49 UT. Its real-time duration is 11 s.\\
{\bf Movie S7}: An animation of the AIA 171, 211, and 94~{\AA} images (Event E3), along with GOES soft X-ray flux in 1-8~\AA\ channel, AIA 304~{\AA} peak counts (extracted from the boxed area in ribbon R3), and Type III radio bursts (245 MHz), during the filament eruption and associated jet (Figure \ref{app-fig3}). The animation runs from 19:01:37 UT to 19:57:25 UT. Its real-time duration is 7 s. \\
{\bf Movie S8}: An animation of the AIA 211~{\AA} intensity and running-difference images (Figure \ref{app-fig4}) during the eruption and the formation of fast EUV waves (Event E3). The animation runs from 17:32:49 UT to 19:57:37 UT. Its real-time duration is 18.1 s.\\
{\bf Movie S9}: An animation of the 3D MHD simulation (current density) of an eruption in the nested-null topology (Figure \ref{fig5}). Its real-time duration is 14.3 s.\\
{\bf Movie S10}: An animation of the 3D MHD simulation (velocity) of an eruption in the nested-null topology (Figure \ref{fig6}). 
 Its real-time duration is 14.3 s.\\

\end{document}